\documentclass{aa}  

\usepackage{graphicx}
\usepackage{txfonts}
\usepackage[colorlinks=true, allcolors=blue]{hyperref}
\usepackage{orcidlink}
\usepackage{adjustbox}
\usepackage{multicol}
\usepackage{multirow}
\usepackage{enumitem} 
\usepackage{cuted} 
\usepackage{xcolor}              

\usepackage{tikz}
\begin{document} 

   \title{Shape and ionization of equatorial matter near compact objects from X-ray polarization reflection signatures}

   \titlerunning{X-ray polarization and equatorial reflection}

\author{J.~Podgorn{\'y} \inst{1}\thanks{E-mail: jakub.podgorny@asu.cas.cz}\orcidlink{0000-0001-5418-291X} 
 } 
  
  \authorrunning{J. Podgorn{\'y}}
  
\institute{Astronomical Institute of the Czech Academy of Sciences, Bo\v{c}n\'{i} II 1401/1, 14100 Praha 4, Czech Republic}

   \date{Received ...; Accepted ...}

 
\abstract
{
Motivated by the success of the Imaging X-ray Polarimetry Explorer (IXPE), providing observational evidence that our Universe is substantially polarized in X-rays, we elucidate what can be inferred about 3D matter structures forming about the equatorial plane of accreting compact objects from 0.1--100 keV linear polarization induced by non-relativistic large-scale reflection. We construct a model of an optically thick elevated axially symmetric reflecting medium with arbitrary ionization profile, representing the known diverse scattering environments: from thick winds and super-Eddington funnel structures formed around black holes and neutron stars, to Compton-thick dusty tori of active galactic nuclei and their broad line regions. We assume a central X-ray power-law source with an isotropic, cosine, and slab-coronal angular distribution, including possible intrinsic polarization. The reprocessing is based on X-ray constant-density local reflection tables produced with a Monte Carlo method combined with detailed non-LTE radiative transfer, although we also show the corresponding examples of fully neutral and fully ionized reflection, including classical (semi-)analytical prescriptions. We conclude that varying ionization has a similarly strong impact on observed polarization as the observer's inclination and the skew and opening angle of the reflector's inner walls, altogether producing up to tens of \% of reflected polarization both parallelly or perpendicularly to the projected system axis, depending on the parameter values combination. After testing 3 different ad-hoc shapes of the reflector: a cone, an elliptical torus, and a bowl, we conclude that while in some configurations, their altered curvature produces more than 30 \% absolute difference in observed total polarization, in others, the adopted shape has a marginal impact. Lastly, we discuss the change of the observed polarization due to relaxing the optically thick assumption on equatorial winds and accreted matter, providing a continuous range of energy-dependent examples between the optically thick and thin scenarios.

   }


   \maketitle
%

\section{Introduction}
\label{sec:intro}

Polarimetry is known for its potential to uncover geometrical details of unresolved objects. In X-rays, the Stokes parameters $I$, $Q$, and $U$ of the reflected radiation are strongly dependent on the ionization state of the reflecting medium due to competing absorption and scattering effects, as well as spectral lines \citep{Matt1993, Poutanen1996b}. Despite other modulations due to magnetic fields, relativistic kinematics, and strong gravity \citep[e.g.,][]{Connors1977,Dovciak2008,Davis2009, Poutanen2023, Barnier2024, Steiner2024}, which are critical in the vicinity of accreting compact objects, basic geometrical and ionization effects remain important when spectro-polarimetrically studying structures orbiting around them at any distance.

In this work, we model geometrically thick equatorial structures, which are optically thick and thus focus light originally emitted close to the central compact object. These are then crudely representing, for example, the cold parsec-scale Compton-thick dusty tori of supermassive black holes in active galactic nuclei (AGN), their nearly neutral sub-parsec-scale broad line regions (BLR), or closer optically thick equatorial outflows and highly-elevated accretion funnels formed due to super-Eddington accretion, which also revolve around accreting stellar-mass black holes and neutron stars inside X-ray binary systems \citep[e.g.,][]{Abramowicz1988,Antonucci1993, Poutanen2007, Neilsen2009, Beckmann2012, Kaaret2017, Giustini2019}. Regardless of the mass of the central object, a hard X-ray power-law radiation is often observed, believed to be a Comptonized thermal emission inside a hot corona located up to a few tens of gravitational radii from the center \citep{Sunyaev1980, Haardt1993b, Poutanen1996}, with possible hard X-ray contributions from the boundary or spreading layers for the case of neutron stars \citep{shakura1988, inogamov1999}. 

We approximate such primary source of X-ray emission as a point source located in the center of an axially symmetric system, assuming it can have a diverse angular distribution of the emitted intensity and polarization, depending on its detailed formation and physical properties \citep{SunyaevTitarchuk1985, Haardt1993, Poutanen1996, Zhang2019, Krawczynski2022b}. We also do not speculate on intricate accretion physics that would lead to a stable geometrical and ionization configuration of the reflecting optically thick structures, but rather revert the problem and ask whether observational X-ray spectro-polarimetry can provide constraints to complex (general-relativistic) magneto-hydro-dynamical simulations \citep[e.g.,][]{Ohsuga2005,Sadowski2014,Jiang2019, Porth2019, Liska2020} or kinetic plasma simulations \citep[e.g.,][]{Parfrey2019,Sironi2020,Nattila2025}. We assume any half-opening angle of the orbiting structure, ranging from extremely collimated ultra-luminous X-ray sources (ULX) where piled-up material beams the central emission up to a few degrees off the main axis \citep{Begelman2006,Kaaret2017,Veledina2024b}, all the way to orbiting material close to the equatorial plane, such as the BLRs in AGNs \citep{Krolik1988,Netzer2015,Gravity2018}. We assume arbitrary skew of the inner walls of the reflector, as well as its curvature: from convex opening structures, considered for photospheres of accretion discs and their winds \citep{Elvis2000,Jiang2014,Conforti2025,Madau2025} and elevated doughnut-shaped (clumpy) dusty tori or BLRs \citep{Wada2002,Schartmann2005, Murphy2009,Chelouche2019}, to concave cusped geometries considered for (super-Eddington) accretion flows \citep{Abramowicz1978, Sadowski2014, Wielgus2016} and bowl-shaped parsec-scale AGN structures \citep{Goad2012,Wada2016,Gravity2024,xrism2024}. 

Although some studies have presented X-ray polarization signatures from 3D reflecting structures near accreting compact objects \citep[e.g.,][]{Ratheesh2021,West2023,Veledina2024, Veledina2024b,Meulen2024}, the detailed role of focusation and ionization profile of the reflector remains unprobed. Here we elaborate on our previous work \cite{Podgorny2024} that introduced a simple {\sc XSPEC} \citep{Arnaud1996} fitting tool called {\tt xsstokes\_torus}. Motivated by the recent discoveries with the Imaging X-ray Polarimetry Explorer \citep[IXPE,][]{Weisskopf2022}  in the 2--8 keV band \citep[see reviews][for the results from the first 2.5 years of the missions operation]{Dovciak2024, Marin2024, Poutanen2024, Ursini2024}, we believe that current and planned X-ray polarimeters, combined with cutting-edge spectroscopy and multi-wavelength information, can shed light on the rich accretion structures. While we leave a construction of an updated fitting tool, along with an example of fitting IXPE data, to a follow-up study, we present in this paper a thorough exploration of the designed parameter space.

In Section \ref{sec:model}, we describe the details of the reflection model under a strict optically thick assumption. In Section \ref{sec:results}, we describe the results for both reflected-only and total emission from this model. Importantly for the interpretation of some of the recent IXPE discoveries and for characterizing limitations of the presented model, we discuss in Section \ref{sec:discussion} the role of the optical thickness of the equatorial obscurers, using additional calculations. We conclude in Section \ref{sec:summary}.

\section{Model}
\label{sec:model}

The presented reflection model is built based on the {\tt torus\_integrator} routine\footnote{Available at \url{https://github.com/jpodgorny/torus_integrator}, including documentation, updated for the extensions presented in this work.} and the corresponding {\tt xsstokes\_torus} model\footnote{Available at \url{https://github.com/jpodgorny/xsstokes_torus}, including documentation.}, presented in \cite{Podgorny2024} for nearly neutral reflection. The polarization prediction is based on numerical integration of local reflection tables\footnote{For direct usage inside {\tt XSPEC} under different flavors, see \url{https://github.com/jpodgorny/stokes_tables}, including documentation.} \citep{Podgorny2022} across a toroidal axially symmetric static structure, mirror symmetric about the equatorial plane. A point-like source of power-law X-ray emission with arbitrary power-law index $\Gamma$ and arbitrary polarization is located in the center of the system, representing the Comptonizing medium. In a Newtonian approximation, neglecting the relativistic effects, we calculate the spectro-polarimetric image that a distant, generally inclined observer would obtain, if each part of the reflector's surface were reflecting once, i.e. without secondary reflections between the inner walls. However, we take into account that each ray from the central source is rather reprocessed inside the walls, than truly reflected. The local reflection tables, fully described in \cite{Podgorny2022}, assume a constant density slab of $n_\textrm{H} = 10^{15} \, \textrm{cm}^{-3}$ and an impinging power-law with 3 independent states of polarization and arbitrary local incident and emission angles. The non-LTE ionization profile of the plane-parallel slab is pre-computed with the {\tt TITAN} code \citep{Dumont2003}. Then, in the same geometrical setup, the spectro-polarimetric properties of the emergent radiation in azimuthal and meridional directions are computed with a Monte Carlo simulator {\tt STOKES} \citep{Goosmann_2007, Marin_2012, Marin_2015, Marin_2018_UV}. In this way, we include the effects of multiple Compton down-scattering on electrons, absorption, and fluorescent and resonant spectral lines, in order to obtain detailed spectral, polarization, and ionization structure insights. We assume solar abundances of elements from \cite{Asplund2005} with $A_\mathrm{Fe} = 1.0$. The photons, which locally escape the slab, are then assumed to emerge from the point where they entered inside the global numerical integration across the reflecting surface. In such plane-parallel optically thick approximation, the curvature of the shapes studied is assumed small on the scale of a photon's mean free path.

In this work, we present several extensions of the numerical scheme. Although we plan to release the corresponding {\sc XSPEC} fitting model in a follow-up paper with an example of usage on the IXPE data, here we discuss the effects of the added model features on the observed X-ray spectra and polarization. We enhance the model with
\begin{itemize}[label=\textbullet]
    \item extending the energy range from 1--100 keV to 0.1--100 keV,
    \item allowing any ionization profile of the reflecting surface, including fully ionized and fully neutral cases,
    \item adding primary radiation to the output for unobscured geometries, hence allowing predictions for the total (= primary + reflected) Stokes parameters $I$, $Q$, and $U$ observed at inclinations above the grazing angle,
    \item allowing additional emission angular distributions to the isotropic central source: a cosine and slab-coronal sources,
    \item generalizing the circular torus geometry of the reflector to an elliptical torus,
    \item adding cone- and bowl-shaped geometries of the reflector.
\end{itemize}
We intend to inspect further assumptions of the model, such as the effects of multiple reflections, special- and general-relativistic modifications, asymmetry or irregularity of the reflector, or the geometrical extension of the source, in the future.

\begin{figure*} 
\centering
\includegraphics[width=1\linewidth]
{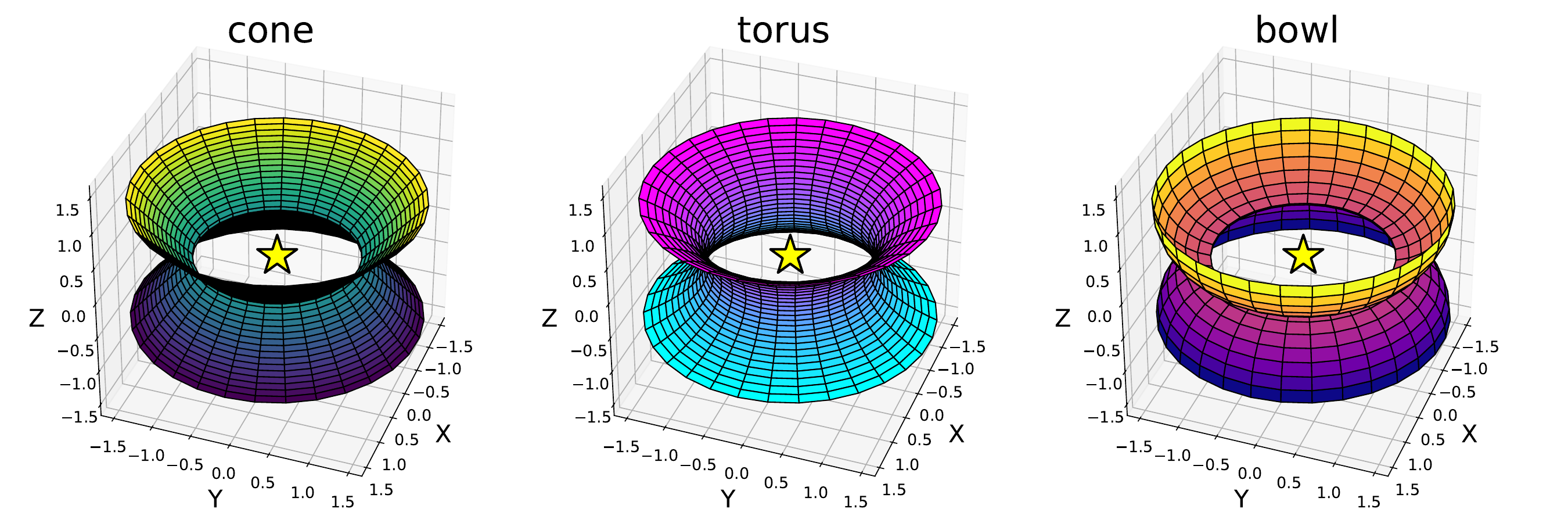}
\caption{The reflecting surface in 3D Cartesian global coordinate system, as viewed by an inclined observer. From left to right we display for the same $\Theta$ and $\rho/\rho_\mathrm{in}$ the cone-shaped, torus-shaped, and bowl-shaped geometries, respectively, that are examined in this work. The yellow star represents the central source of emission.} \label{fig:sketch3D}
\end{figure*}

\begin{figure} 
\centering
\includegraphics[width=1\linewidth, trim={3.2cm 4.9cm 7.8cm 3.1cm}, clip]
{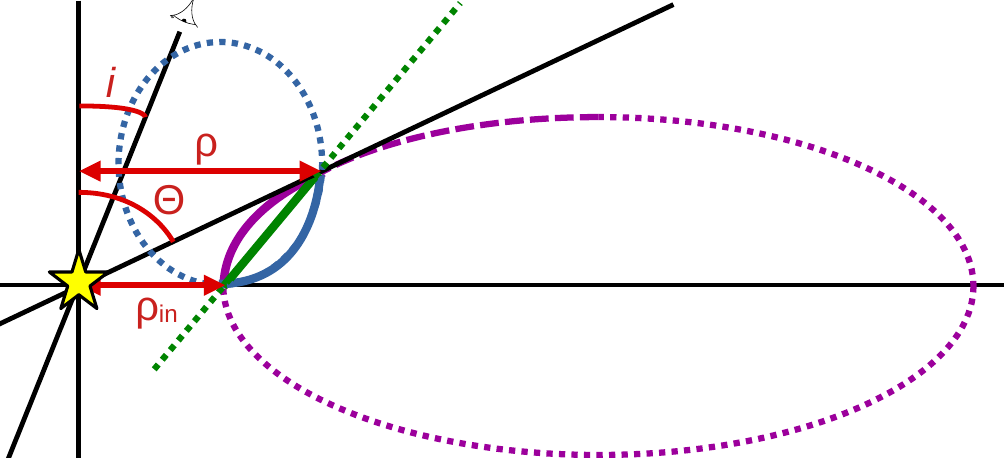}
\caption{The parametrization of the cone-shaped (green), torus-shaped (purple), and bowl-shaped (blue) reflectors. The sketch is in the meridional plane, and the reflecting structure is further rotated around the main axis and remains axially symmetric. We highlight in solid lines only the upper half of the reflecting surface above the equatorial plane in each geometry, although the model allows for taking into account reflection from the bottom half-space of a mirror-symmetric surface to the equator. In the dashed purple line, we show the part of the inner walls of the torus, which may self-obscure some other reflecting part of the torus for a highly inclined observer, but is not directly illuminated by the source due to self-shadowing. We show in the dotted purple line the ellipse defining the torus-shaped reflector in both half-spaces from the equator. Similarly, we show in the dotted green line the corresponding straight edge of an off-centered geometrical cone in the meridional plane that defines the cone-shaped reflector in the upper half-space. Similarly, we also show in the dotted blue line the ellipse defining the bowl-shaped reflector in the upper half-space. The observer is inclined at an inclination $i$. Each reflecting geometry is fully described by its half-opening angle $\Theta$, the inner radius $\rho_\mathrm{in}$, and the maximal illuminated radius $\rho$. When $i > \Theta$, the primary source is obscured for the observer.} \label{fig:geometry_sketch}
\end{figure}

We choose to present three distinct ad-hoc shapes of the reflecting inner walls of the thick equatorial obscurers. The cone, the elliptical torus, and the bowl shapes represent a straight, convex, and concave examples of the wall shape with respect to the main axis in the meridional plane. They are schematically shown from the observer's point of view in Fig. \ref{fig:sketch3D} and parametrized in Fig. \ref{fig:geometry_sketch} inside the meridional plane. For each case, the surface is rotationally symmetric around the principal axis and mirror symmetric with respect to the equatorial plane, although the sketch shows only one quadrant for greater visual clarity. Similarly to the original model in \cite{Podgorny2024}, we define a switch $B = \{1,0\}$, which states whether we do, or do not, take the reflection from a bottom half-space below the equator into account. The observer is inclined under angle $i \in [0^\circ;90^\circ)$ with respect to the main axis, which defines the ``top'' half-space with respect to the equator, where the observer is located. Any model reflector is defined by its half-opening angle $\Theta \in (0^\circ;90^\circ)$ with respect to the main axis. The minimal distance of the reflector from the center is for each case in the equatorial plane, defined as $\rho_\mathrm{in}$. The maximal radius of the directly illuminated section of the reflector's inner walls is defined as $\rho$ and corresponds to the tangential point between the cone dissected by $\Theta$ and the reflector. The ratio $\rho/\rho_\mathrm{in}$ then defines the skew of the inner walls. The height of the reflecting section of the obscurer above the equatorial plane is then $h = \rho /\tan{\Theta}$.

The cone-shaped reflector is in each half-space defined by an off-center geometrical cone, i.e. with a straight surface boundary in the meridional plane and with a maximal height $h$. The torus-shaped reflector is an elliptical torus, defined by a geometrical ellipse in the meridional cut, centered at the equatorial plane. The parameters of such a general ellipse are fully described by the $\rho_\textrm{in}$ distance, setting the closest point of the ellipse to the center, and by $\rho$ and $\Theta$, setting the tangential point of the ellipse to the cone dissected by $\Theta$. The special case of a circular torus profile, presented for the original model in \cite{Podgorny2024}, is re-obtained when setting
\begin{equation}\label{rho_circular}
    \rho = \rho_\mathrm{c} \equiv \frac{\rho_\mathrm{in}}{(1-\cos{\Theta})(1+\cot^2{\Theta})} \, \, .
\end{equation}
Otherwise, $\rho$ is a free parameter of the model. The bowl-shaped reflector is in each half-space defined by a quarter of a meridional ellipse, which is tangential to the equator, and its center is at the inner radius $\rho_\mathrm{in}$ and at the height $h = \rho /\tan{\Theta}$. We select the quarter dissected by the axes of the ellipse, which is closer to the equator and further from the rotational axis. Note that when following such definitions, the inner walls of the reflecting cone and bowl are entirely illuminated by the central source and a fraction of a reflecting surface can by viewed by an observer inclined at any inclination $i$. The elliptical torus, if considered in its full extent, generally self-shields its upper or lower surface from the central source, hence, its actual height is not equal to the height $h$ of its illuminated surface. While in Fig. \ref{fig:sketch3D} we show only the illuminated part of its surface, for the case of the elliptical torus high $i$ and/or low $\Theta$ and/or high $\rho/\rho_\mathrm{in}$ may result in complete obscuration of the illuminated surface. The allowed range of skew of the above-defined elliptical torus is $\rho/\rho_\mathrm{in} \in (1;2)$, while for the cone- and bowl-shaped reflectors it is $\rho/\rho_\mathrm{in} \in [1;\infty)$ and $\rho/\rho_\mathrm{in} \in (1;\infty)$, respectively. We refer to Appendix \ref{computations} for further details on the geometrical implementation.

Defined in this way, the elliptical torus is an approximation to any convex reflecting shape, whose normal is perpendicular to the main axis of symmetry at the inner radius, forming a smooth function at $\rho_\mathrm{in}$, and whose edge above the equatorial plane is reaching the cone dissected by $\Theta$ with an identical normal vector (up to a sign) to the dissecting cone at the tangential point. The bowl is, on the contrary, an approximation to any concave focusing by the reflector, whose normal vector is perpendicular to the main axis at height $h$, but converging to a parallel normal vector to the main axis near $\rho_\mathrm{in}$. The inner radius is then forming a sharp spike, leading to a cusped structure. The cone is an approximation to a flat reflector, i.e. a somewhat intermediate case with a generally, but constantly inclined normal vector with respect to the axis across its half-plane surface, resulting in a continuous but not smooth function at $\rho_\mathrm{in}$.

We envisage that such 3 designed example shapes well represent the effects of changing focusing by different shape of the inner walls of a physical reflector that may exhibit more complex geometries. We do not compute the results for any intermediate configurations between these three, nor for different prescription functions of the reflecting concave, convex, or straight surfaces. We anticipate that should the cone- or bowl-shaped structures further extend above their directly illuminated surfaces via inflection point and convex profile at larger radii and higher heights, the additional self-obscuration effects would be similar to those of the torus. Should a combination of a concave and a convex profile exist within the directly illuminated section of the inner walls in real systems, we presume a mixed polarization output between the torus and the bowl results. Clumpiness of the elevated accreting structures should generally depolarize the output, as the net result will comprise more diverse single-scattering angles compared to a smooth profile reflection. But apart from the expected dilution of the observed polarization fraction, the net linear polarization angle and the geometrical effects of different shapes, skew, inclination, opening angle, or emission properties of the source are likely to qualitatively remain, as the local reflection will average to a dominant scattering angle set by an underlying large-scale smooth shape of the reflector.

The primary source of radiation may exhibit diverse angular distribution of radiation power, as well as its polarization properties. Both intrinsic luminosity $L$ and the incident linear polarization degree $p_0$ can be characterized with respect to $\mu \in [0;1]$, which is the cosine of the emitting angle measured from the main axis in both half-spaces dissected by the equatorial plane. We assume that the angular distribution of the intrinsic flux and polarization is not itself energy-dependent. Then $L(\mu) = L_0a_0(\mu)$, where $L_0$ is the mean (angle-integrated) luminosity and $\int_0^1a_0(\mu)\textrm{d}\mu = 1$. In the examples shown in this work, we consider three cases. An unpolarized isotropic source with $a_0(\mu) = 1$ and $p_0(\mu) = 0$, loosely approximating, e.g., Comptonized emission from the neutron-star spreading layer. An unpolarized cosine source with $a_0(\mu) = 2\mu$ and $p_0(\mu) = 0$, representing disc-like emitters, such as Comptonized emission in the boundary layer of neutron stars. And an approximation of a static slab-like Comptonization corona at 4 keV, provided by model B in \cite{Poutanen2023} with Thomson optical depth $\tau_\mathrm{T} = 1$ and electron temperature $kT_\mathrm{e} = 100$ keV, recently characterized in \cite{Nitindala2025} (Eq. 7) as
\begin{equation}\label{slab_prescription}
\begin{aligned}
    a_0(\mu) &= 1.73 \mu \frac{1+5.3\mu - 0.2\mu^2}{1+1.3\mu+4.4\mu^2} \,\, , \\
    p_0(\mu) &= 0.064(1 - \mu)\frac{1+16.3\mu + 6.2\mu^2}{1+8.2\mu-2.1\mu^2} \, \, .
\end{aligned}
\end{equation}
The corresponding linear polarization angle is for all $\mu$ aligned with the projected system axis to the polarization plane. We note, however, that nearly the same emission polarization profile as for this case of slab-like corona, including the magnitude and the sign of polarization, was obtained in \cite{Podgorny2023a} (Fig. 7) for the lamp-post corona geometry with low height of the corona and low black-hole spin. In this case, the high polarization nearly parallel with the main axis is induced by relativistic reflection from the accretion disc, while the corona is assumed to be unpolarized. A distinction between the polar and equatorial coronal geometries may be inferred for specific sources from both X-ray polarization and spectroscopic information simultaneously \citep{Krawczynski2022}, while for others, it remains ambiguous (Kammoun et al., in prep.). Therefore, the latter prescription of incident radiation may serve as a general estimate from disc-corona modelling, inline with the IXPE data of Comptonization-dominated sources \citep{Dovciak2024, Marin2024, Poutanen2024, Ursini2024}. Although we plan to include arbitrary incident polarization to the updated {\sc XSPEC} model, its effects on the resulting polarization compared to unpolarized source were already discussed in detail in Appendix D of \cite{Podgorny2024}. The discussion remains qualitatively valid also for the added shape, anisotropy, and ionization model features in this work, and we will only compare the above-mentioned 3 examples. Similarly, we will show the results only for $\Gamma = 2$ and refer to \cite{Podgorny2024} for notes on the effects of changing $\Gamma$.

The torus reflection model presented in \cite{Podgorny2024} took into account only nearly neutral reflection, integrating the local reflection tables in their least ionized value of the ionization parameter \citep{Tarter1969} $\xi = 4\pi F_\mathrm{inc}/n_\mathrm{H} = 5 \, \textrm{erg} \cdot \textrm{cm} \cdot \textrm{s}^{-1}$, where $F_\mathrm{inc}$ is the total local flux illuminating the surface. In this work, we use the full ionization range of the local reflection tables and set an arbitrary power-law ionization surface profile:
\begin{equation}\label{xidef}
    \xi(r,\mu) = \xi_0 \, a_0(\mu) \,\, \mu_\textrm{i}(r) \, \left( \frac{\rho_\mathrm{in}}{r} \right)^\beta \,\, ,
\end{equation}
where $\xi_0$ is the mean ionization parameter at the $\rho_\mathrm{in}$ distance from the source, $r$ is the distance of the surface point from the center, $\beta$ is the illumination power-law index, and $\mu_\mathrm{i}$ is the cosine of the local incident inclination angle with respect to the surface normal. We choose to parameterize the model with $\xi_0$, as for pure reflection, it allows to estimate both intrinsic flux and reflector's density impact simultaneously without the need to specify one or another. For the latter we do not have spectro-polarimetric reflection tables pre-computed for different true densities than $n_\mathrm{H} = 10^{15} \, \textrm{cm}^{-3}$, always constant in the local normal direction inside the reflector, locally approximated as a plane-parallel slab. But we estimate that the true density change is expected to affect mostly the thermal peak in the reflection spectra below $\sim 2$ keV for $n_\mathrm{H} > 10^{15} \, \textrm{cm}^{-3}$ \citep{Garcia2016}. Hence, above $\sim 2$ keV, changing $\xi_0$ and $\beta$ can be considered as changing the reflector's surface density profile for an assumed intrinsic luminosity. 

A constant-density reflection profile across its illuminated surface is obtained for $\beta = 2$, because the flux diminishes as $\sim 1/r^2$. The X-ray luminosity of the source between the adopted sharp power-law cutoffs at $E_\mathrm{min} = 10^{-1.1}$ keV and $E_\mathrm{max} = 10^{2.4}$ keV can then be obtained as
\begin{equation}
    L_\mathrm{X} = \xi_0 \, n_\mathrm{H,0} \, \rho_\mathrm{in}^2 \, \, ,
\end{equation}
which is exact when assuming 
\begin{equation}
   n_\mathrm{H,loc}(r) = n_\mathrm{H,0} = 10^{15} \, \textrm{cm}^{-3} = \textrm{const.}
\end{equation}
across the surface. Otherwise, it is an estimate of the total luminosity for a density profile of 
\begin{equation}\label{dendef}
    n_\mathrm{H,loc}(r) =  n_\mathrm{H,0}\, \left(  \frac{r}{\rho_\mathrm{in}}\right)^{\beta - 2} \, \, ,
\end{equation}
which holds reasonably well for average $n_\mathrm{H,loc}(r) \lesssim 10^{15} \, \textrm{cm}^{-3}$ and/or energies $E \gtrsim 2$ keV. Only if we wish to add self-consistently the primary radiation to the result (when examining $i < \Theta$), we need to additionally specify either $L_\mathrm{X}/\rho_\mathrm{in}^2$ or $n_\mathrm{H,0}$ for a given $\xi_0$. For observable polarization properties of the system, it is not necessary to specify physical units of distance; only relative distances inside the system matter. We refer to Appendix \ref{normalization} for further details on the normalization of the primary and reflected flux.

For the fully ionized examples, we use the Chandrasekhar's prescription for diffuse reflection \citep[][Section 70.3]{Chandrasekhar1960}, assuming a 100\% albedo for the locally incident radiation. For the fully neutral examples, we plot only a polarization prediction using the single-scattering Thomson law with intensity angular redistribution according to \citet[][Section 47.1]{Chandrasekhar1960}, or we use a newly created purely neutral version\footnote{For direct usage of the neutral reflection tables inside {\tt XSPEC}, see also \url{https://github.com/jpodgorny/stokes_tables}, including documentation.} of the local reflection tables described in \citet{Podgorny2022}. The latter has been computed as part of this work in an identical setup to the partially ionized reflection tables and with the same version of the Monte Carlo code {\tt STOKES}. For showing the spectra of any of these special cases in the following section, we choose a particular $\xi_0$, $\rho_\mathrm{in}$, distance to the source $D$, and normalize according to Appendix \ref{normalization} for a given $n_\mathrm{H,0}=10^{15} \, \textrm{cm}^{-3}$.

\section{Results}
\label{sec:results}

\begin{figure*} 
\centering
\begin{tikzpicture}[
x=1pt, y=1pt,
inner sep=0pt,
outer sep=0pt]
\node[anchor=south west] (base) at (0,0)
  {\includegraphics[width=1\linewidth, trim={3.7cm 1.7cm 4.4cm 3.9cm}, clip]
{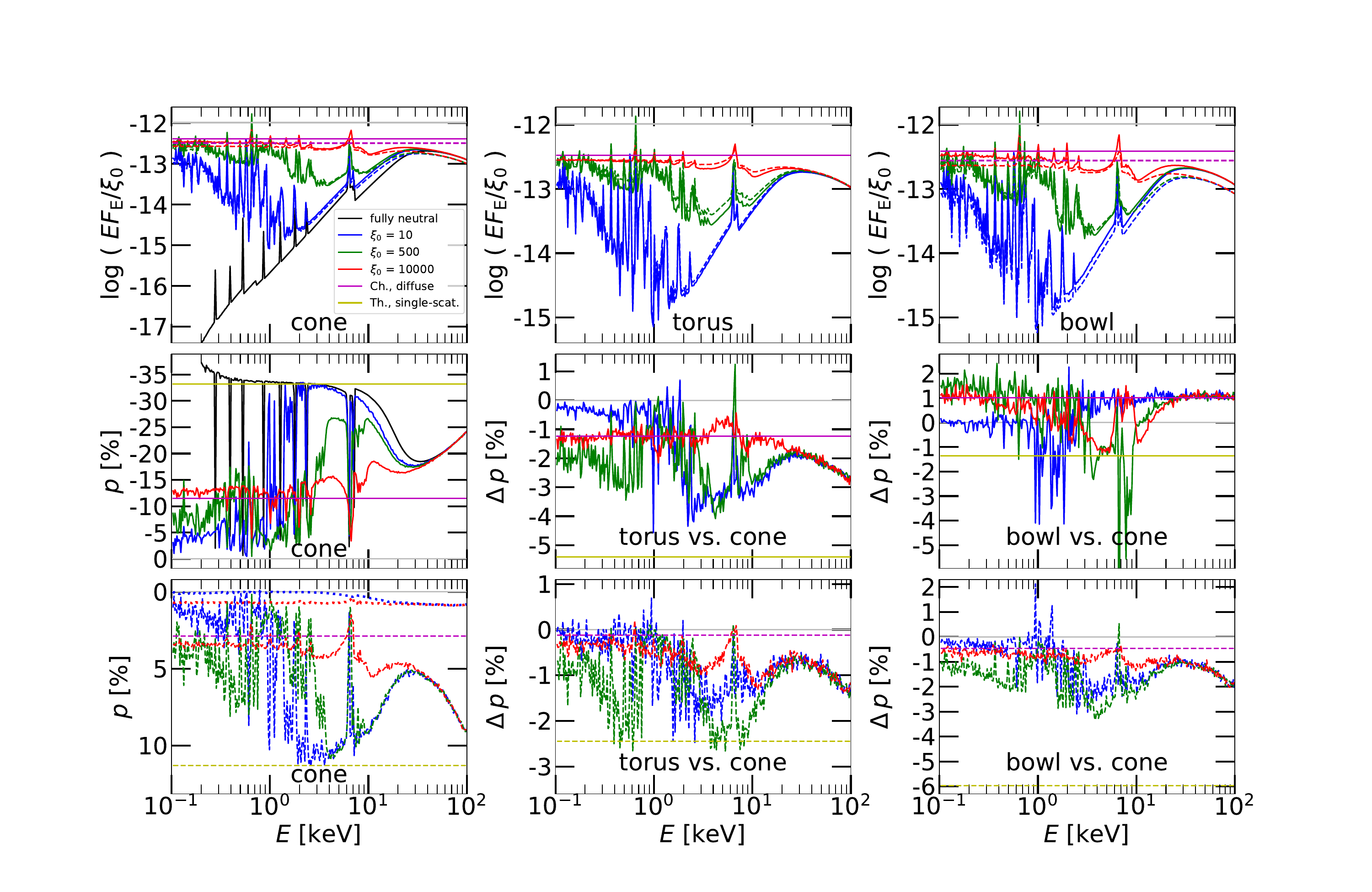}};
\node[anchor=south west] at (137,185)
  {\includegraphics[width=0.035\linewidth]{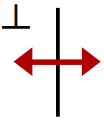}};
\node[anchor=south west] at (135.5,39)
  {\includegraphics[width=0.027\linewidth]{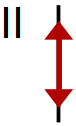}};
\end{tikzpicture}
\caption{\textbf{Top row:} the spectra for a distant observer at $D = 1$ kpc and $\rho_\mathrm{in} = 10^9$ cm. We show the logarithm of $EF_\mathrm{E}/\xi_0$ in $\textrm{cm}^{-3}$ to compensate for the slope given by the power-law index $\Gamma = 2$ and to level the amplitude, given to a large extent by the ionization parameter $\xi_0$. In the color code we show the reflected-only spectra for $\xi_0 = 10 \, \, \textrm{erg} \cdot \textrm{cm} \cdot \textrm{s}^{-1}$ (blue), $\xi_0 = 500 \, \, \textrm{erg} \cdot \textrm{cm} \cdot \textrm{s}^{-1}$ (green), $\xi_0 = 10\,000 \, \, \textrm{erg} \cdot \textrm{cm} \cdot \textrm{s}^{-1}$ (red), and for the Chandrasekhar's formulae for diffuse reflection with 100\% albedo (magenta). The primary radiation is shown in gray. The observer is inclined at $i = 51^\circ$ and the half-opening angles are $\Theta = 40^\circ$ (solid lines) and $\Theta = 70^\circ$ (dashed lines). We show the reflection from a cone (left), torus (center), and bowl (right), all for $\rho = \rho_\mathrm{c}$, $B = 1$, and unpolarized isotropic irradiation with $\beta = 2$. Specifically for the cone geometry and $\Theta = 40^\circ$, we show in black solid line the corresponding spectra in the same configuration, but integrating the fully neutral reflection tables obtained with the {\tt STOKES} code. \textbf{Middle and bottom row:} the corresponding reflected-only polarization degree, $p$, versus energy. For clarity, we show the results for different opening angles $\Theta = 40^\circ$ (solid lines) and $\Theta = 70^\circ$ (dashed lines) in separate rows of panels. In the color-code, we provide in addition the results for Thomson single-scattering approximation (yellow). For the cone geometry and $\Theta = 70^\circ$ (bottom left), we show in addition the total polarization degree for $\xi_0 = 10 \, \, \textrm{erg} \cdot \textrm{cm} \cdot \textrm{s}^{-1}$ (blue dotted lines) and $\xi_0 = 10\,000 \, \, \textrm{erg} \cdot \textrm{cm} \cdot \textrm{s}^{-1}$ (red dotted lines). For the torus (center) and bowl (right) geometries, we show the polarization degree difference $\Delta p$ in \%, which is the polarization degree $p$ of the reflecting cone subtracted from the polarization degree $p$ of the reflecting torus or bowl, respectively, in identical configurations. This is to display examples of the impact of changing curvature of the inner walls of the reflector on the resulting polarization.} \label{fig:energy_dep}
\end{figure*}

\begin{figure*} 
\centering
\includegraphics[width=1\linewidth, trim={12cm 3.3cm 10.5cm 1.8cm}, clip]
{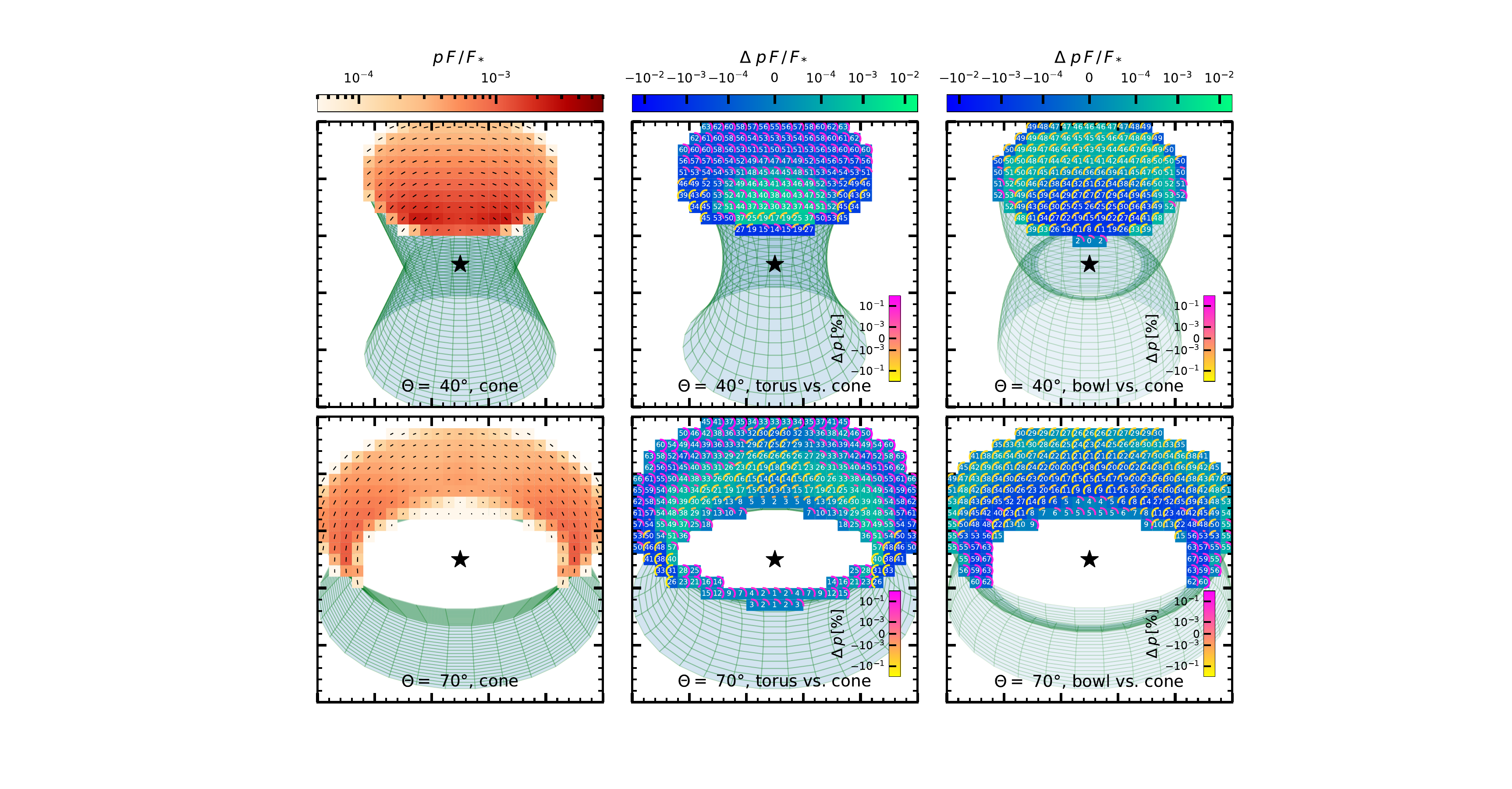}
\caption{Images of the reflecting walls of a cone (left), a torus (center), and a bowl (right) with $\xi_0 = 500 \, \, \textrm{erg} \cdot \textrm{cm} \cdot \textrm{s}^{-1}$ and half-opening angles $\Theta = 40^\circ$ (top) and $\Theta = 70^\circ$ (bottom). Each image contains 25x25 pixels with spectro-polarimetric information integrated in the 3.5--6 keV band, if subtending the observable reflecting area. We use the color code for the background of each reflecting pixel to emphasize the polarized flux, $pF/F_*$, where $F_*$ is the flux of the primary source in observer's direction, for the cone; and the polarized flux difference, $\Delta pF/F_*$, for the torus and bowl, which is the polarized flux, $pF/F_*$, of the reflecting cone subtracted from the polarized flux, $pF/F_*$, of the reflecting torus or bowl in the same pixel for identical configurations. For the cone, each reflecting pixel contains a polarization bar, whose length is proportional to the observed polarization degree from that pixel and whose tilt from the vertical direction is corresponding to its polarization angle. For the torus and bowl, each reflecting pixel contains a number, which is the observed polarization degree in \% from that pixel, and a separately color-coded arc, whose length, direction and color are altogether highlighting the polarization degree difference $\Delta p$ in \%, which is the polarization degree $p$ of the reflecting cone subtracted from the polarization degree $p$ of the reflecting torus or bowl, respectively, in the same pixel for identical configurations. All other parameter values are the same as in Fig. \ref{fig:energy_dep}.} \label{fig:imaging}
\end{figure*}

\begin{figure*} 
\centering
\begin{tikzpicture}[
x=1pt, y=1pt,
inner sep=0pt,
outer sep=0pt]
\node[anchor=south west] (base) at (0,0)
  {\includegraphics[width=1\linewidth]{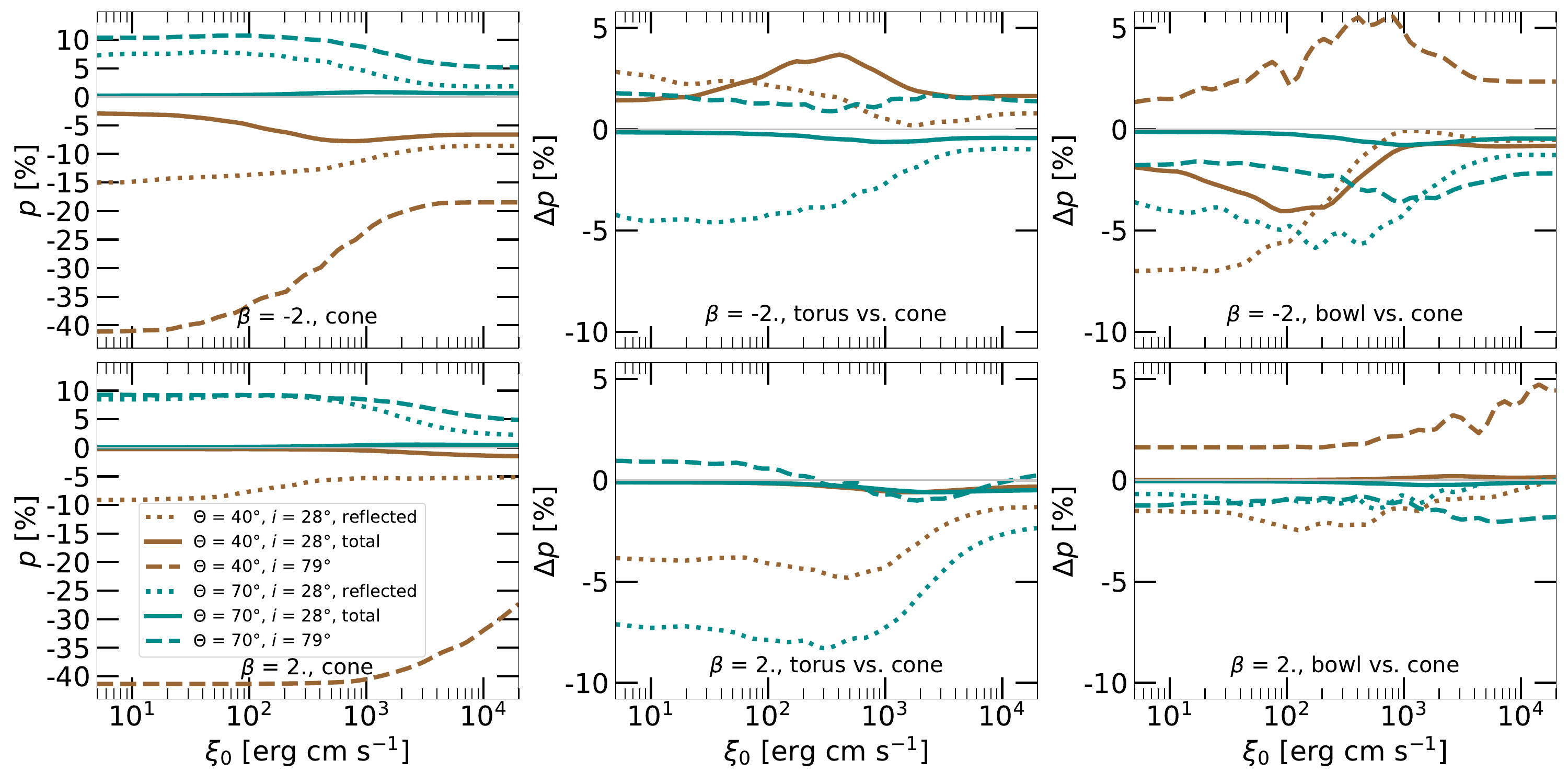}};
\node[anchor=south west] at (2,235)
  {\includegraphics[width=0.018\linewidth]{para.png}};
\node[anchor=south west] at (2,162)
  {\includegraphics[width=0.024\linewidth]{perp.png}};
\node[anchor=south west] at (2,120)
  {\includegraphics[width=0.018\linewidth]{para.png}};
\node[anchor=south west] at (2,45)
  {\includegraphics[width=0.024\linewidth]{perp.png}};
\end{tikzpicture}
\caption{The 3.5--6 keV averaged polarization degree versus the equatorial ionization parameter, $\xi_0$, for two different surface ionization profiles with $\beta = -2$ (top) and $\beta = 2$ (bottom). For the cone geometry (left), we show the reflected-only polarization degree, $p$ in \%. For the torus (center) and bowl (right) geometries, we show the polarization degree difference, $\Delta p$ in \%, which is the polarization degree $p$ of the reflecting cone subtracted from the polarization degree $p$ of the reflecting torus or bowl, respectively, in identical configurations. The results are plotted for $\Theta = 40^\circ$ (brown) and $\Theta = 70^\circ$ (turquoise), and for two different inclinations: $i = 79^\circ$, where reflected radiation is the only observable component (dashed lines), and $i = 28^\circ$, where we show the reflected-only polarization (dotted lines) and the total observed polarization (solid lines). We show the case of $\rho = \rho_\mathrm{c}$, $B = 1$, and unpolarized isotropic irradiation. For the torus, the configuration with $\Theta = 40^\circ$ and $i = 79^\circ$ (brown dashed lines) is already under full eclipse, hence it is not displayed.} \label{fig:xi_dep}
\end{figure*}

\begin{figure*} 
\centering
\begin{tikzpicture}[
x=1pt, y=1pt,
inner sep=0pt,
outer sep=0pt]
\node[anchor=south west] (base) at (0,0)
  {\includegraphics[width=1\linewidth, trim={0.3cm 1.5cm 0.7cm 3.7cm}, clip]
{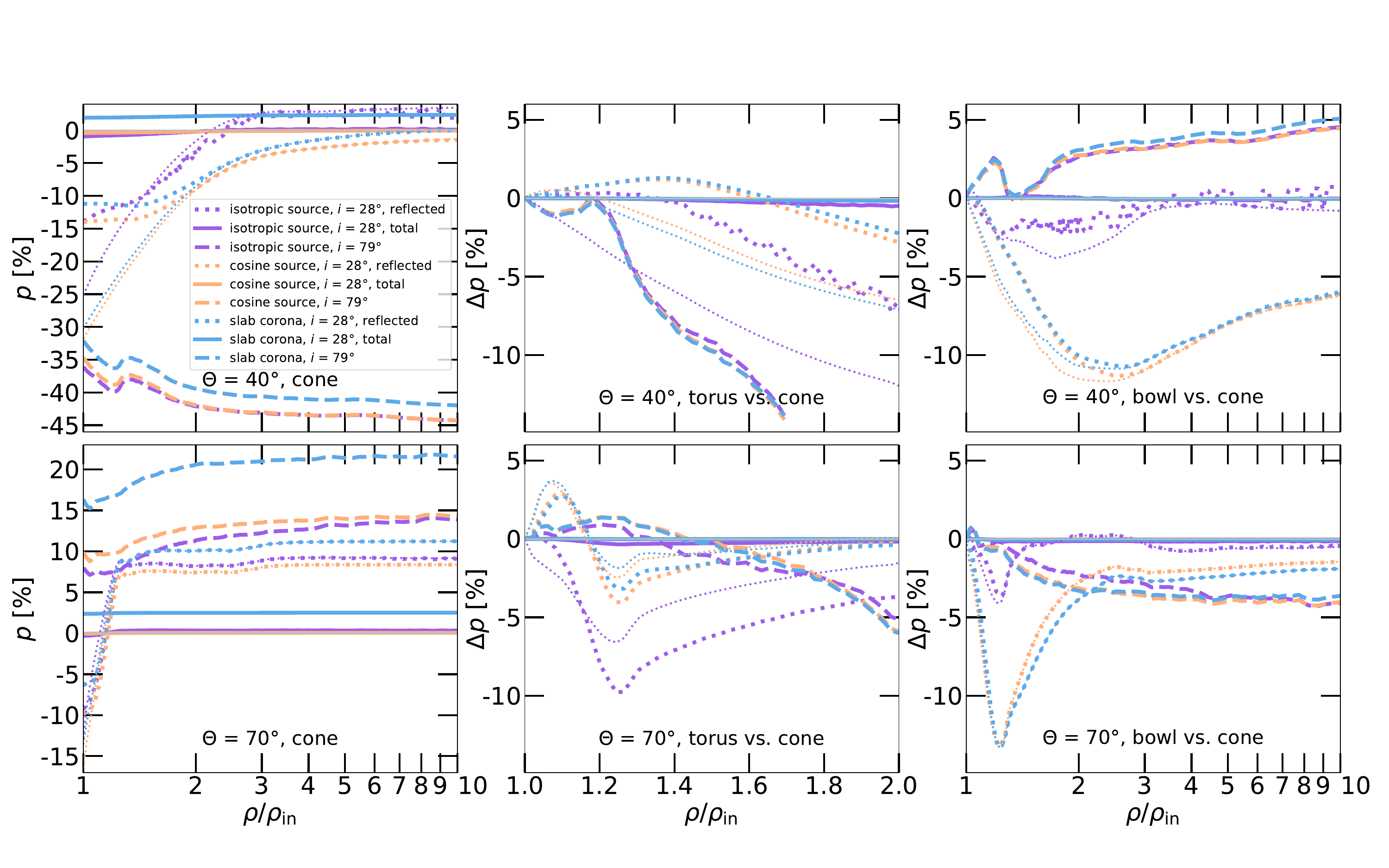}};
\node[anchor=south west] at (0.5,265)
  {\includegraphics[width=0.018\linewidth]{para.png}};
\node[anchor=south west] at (0,165)
  {\includegraphics[width=0.025\linewidth]{perp.png}};
\node[anchor=south west] at (0.5,112)
  {\includegraphics[width=0.018\linewidth]{para.png}};
\node[anchor=south west] at (1,43)
  {\includegraphics[width=0.025\linewidth]{perp.png}};
\end{tikzpicture}
\caption{The 3.5--6 keV averaged polarization degree versus the skew of the inner walls, $\rho/\rho_\mathrm{in}$, for two different half-opening angles $\Theta = 40^\circ$ (top) and $\Theta = 70^\circ$ (bottom). For the cone geometry (left), we show the reflected-only polarization degree, $p$ in \%. For the torus (center) and bowl (right) geometries, we show the polarization degree difference, $\Delta p$ in \%, which is the polarization degree $p$ of the reflecting cone subtracted from the polarization degree $p$ of the reflecting torus or bowl, respectively, in identical configurations. The results are plotted for the case of unpolarized isotropic source (purple), unpolarized cosine source (orange), and the slab corona example (blue), and for two different inclinations: $i = 79^\circ$, where reflected radiation is the only observable component (dashed lines), and $i = 28^\circ$, where we show the reflected-only polarization (dotted lines) and the total observed polarization (solid lines). In regular thick lines, we show examples for $B = 1$, but for the case of low inclination $i = 28^\circ$ and reflected-only emission (dotted lined) we additionally display in thin lines the examples for $B = 0$. The reflection from the bottom half-space of the structures plays the largest role for small $\rho/\rho_\mathrm{in}$. The results are displayed for $\xi_0 = 500 \, \, \textrm{erg} \cdot \textrm{cm} \cdot \textrm{s}^{-1}$ with $\beta = 2$. For the case of torus geometry, $\Theta = 40^\circ$ and $i = 79^\circ$, i.e. the dashed lines in the top central panel, the configurations with $\rho/\rho_\mathrm{in} \gtrsim 1.7$ are already under full eclipse, hence they are not displayed.} \label{fig:rho_dep}
\end{figure*}

\begin{figure*} 
\centering
\begin{tikzpicture}[
x=1pt, y=1pt,
inner sep=0pt,
outer sep=0pt]
\node[anchor=south west] (base) at (0,0)
  {\includegraphics[width=1\linewidth, trim={3.1cm 0.5cm 4.4cm 0.5cm}, clip]
{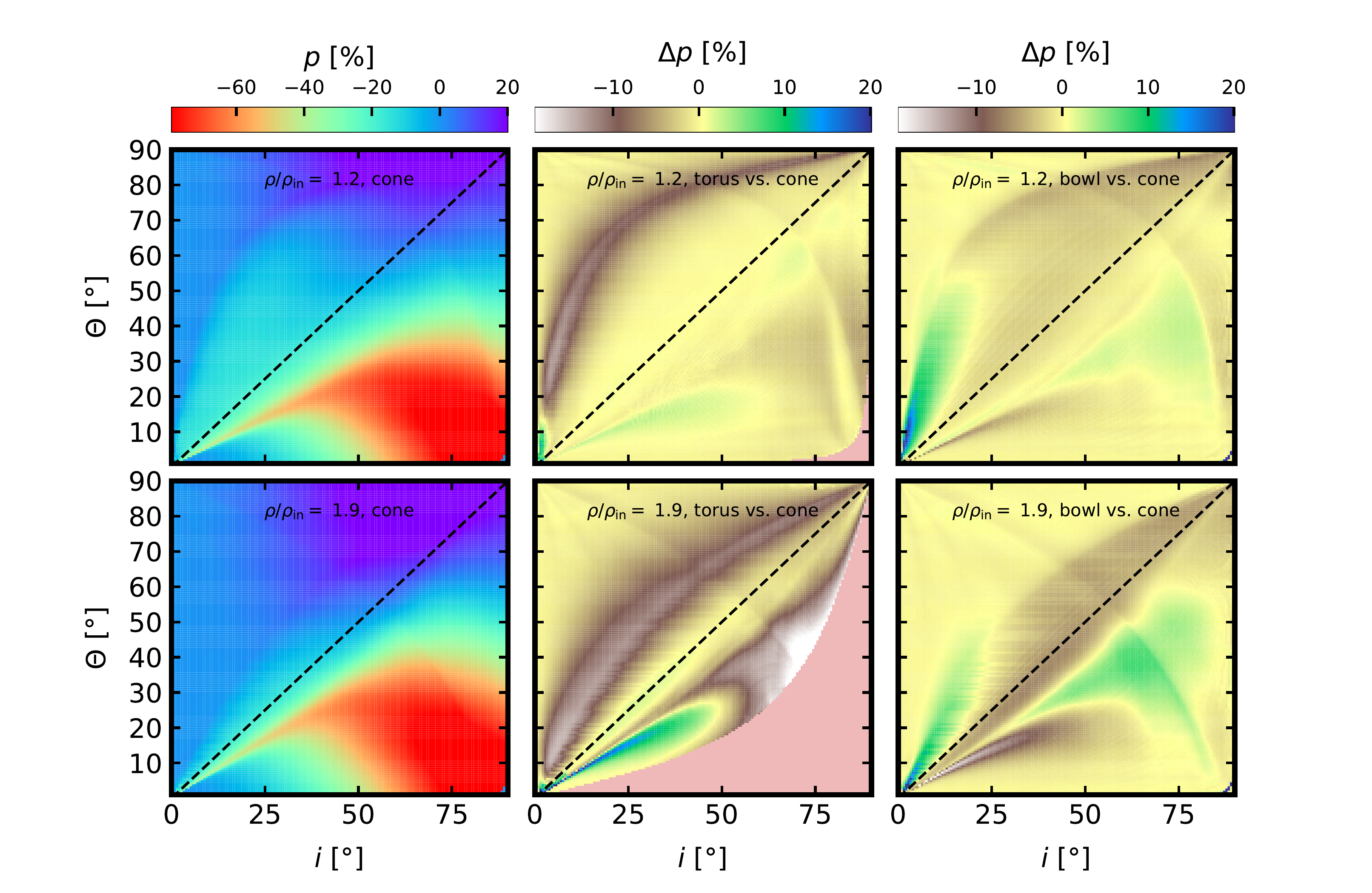}};
\node[anchor=south west] at (168,356)
  {\includegraphics[width=0.018\linewidth]{para.png}};
\node[anchor=south west] at (45,356)
  {\includegraphics[width=0.025\linewidth]{perp.png}};
\end{tikzpicture}
\caption{The 3.5--6 keV averaged polarization degree versus observer's inclination, $i$, and reflector's half-opening angle, $\Theta$, for $\rho/\rho_\mathrm{in} = 1.2$ (top) and $\rho/\rho_\mathrm{in} = 1.9$ (bottom). For the cone geometry (left), we show the reflected-only polarization degree, $p$ in \%. For the torus (center) and bowl (right) geometries, we show the polarization degree difference $\Delta p$ in \%, which is the polarization degree $p$ of the reflecting cone subtracted from the polarization degree $p$ of the reflecting torus or bowl, respectively, in identical configurations. The results are plotted for $B = 1$, $\xi_0 = 500 \, \, \textrm{erg} \cdot \textrm{cm} \cdot \textrm{s}^{-1}$, $\beta = 2$, and isotropic unpolarized source. The dashed diagonal line corresponds to the grazing-angle inclination. The pink-shaded areas in the central panels for high $i$ and low $\Theta$ of the torus correspond to the configurations where the entire illuminated part of the torus is obscured.} \label{fig:heatmaps}
\end{figure*}

The reflection polarization model parameters are $i$, $\Theta$, $\rho/\rho_\mathrm{in}$, $B$, $\xi_0$, and $\beta$, which we will test in the aforementioned 3 geometric shapes and the 3 cases of X-ray emitters inside the accreting system. We will denote the emergent linear polarization degree as $p$. Because of symmetry and the adopted conditions, the net linear polarization angle will always be either parallel or perpendicular to the projected main axis. Therefore, we will denote the parallel and perpendicular orientations as positive and negative $p$, respectively. For greater clarity, we indicate at each set of polarization panels the corresponding polarization angle direction with respect to the projected system axis.
\\
\\
In Fig. \ref{fig:energy_dep} we show for various geometrical, as well as ionization prescriptions, the observed unabsorbed energy flux, $EF_\mathrm{E}/\xi_0$, corrected for the energy slope of the primary radiation and for the changing spectral amplitude effects due to ionization. We also provide the corresponding energy dependence of reflection-induced and total polarization. Although geometry has a limited impact on the spectra, polarization is highly sensitive to the related parameters. Not only from the two provided examples of $\Theta = 40^\circ$ and $\Theta = 70^\circ$ for intermediate inclination $i = 51^\circ$, but from a larger parameter space explored, we confirm the findings of \citet{Podgorny2023b, Podgorny2024, Meulen2024} that for large/small half-opening angles one receives X-ray polarization parallel/perpendicular to the main axis due to predominantly equatorial/meridional single-scattering events, resulting in polarization perpendicular to the scattering plane. We refer to the same publications for the discussion of opening-angle and inclination effects, and to the applications in \citet{Podgorny2024b}, which included a discussion of polar scatterings, leading to the polar or highly-elevated reflection interpretation of the IXPE observations of the Circinus Galaxy AGN, to some extent supported by a simultaneous study \citet{Tanimoto2023}. For an optically thick reflector, the net sign of polarization is determined by the competing polarized flux contributions from the observed left and right sides (leading to parallel polarization angle with the axis), and from the front and back sides (leading to perpendicular polarization angle with the axis). Thus, the resulting sign of polarization is primarily determined by geometrical parameters of the system, while the reflector's ionization profile predominantly affects the net polarization magnitude.

Ionization strongly modulates both spectra and polarization with energy in the X-rays due to the competing scattering, absorption and spectral line effects. We confirm that the integrated spectral shape across the reflecting medium with changing ionization follows that from the adopted local reflection tables \citep{Podgorny2022}, which is a well-known result for power-law reflection from a constant-density slab \citep[e.g.,][]{Fabian2000}. We obtain about 40\% (depending on exact geometry) of the primary flux in reflection, assuming a fully ionized surface with a 100\% albedo. The obtained polarization dependence on energy is locally given by depolarization in fluorescent spectral lines, depolarization in the Compton hump due to multiple scatterings, and enhancement of polarization via absorption, which were all discussed in detail in \citet{Podgorny2022, Podgorny2023a}. Here we newly compare the partially ionized results with the fully neutral and fully ionized limits. We see that regardless of the geometry, the highly ionized integrated reflection tables converge to the Chandrasekhar's result for diffuse reflection, which sets the energy-independent (and relatively low) levels of reflected polarization degree undisturbed by the spectral lines. The high-energy tail in increased polarization, observed in all results from the underlying Monte Carlo simulations, is due to Compton down-scattering. Even the purely neutral integrated Monte Carlo tables (black lines) include an ``inverted'' Compton hump above $\sim 10 \, \textrm{keV}$ due to competing effects of multiple scatterings and decreasing absorption opacity towards hard X-ray energies. The spectral lines and related depolarization is the strongest for the most neutral scenario. The continuum below $\sim 5 \, \textrm{keV}$ is, however, for the purely neutral Monte Carlo results rather energy independent. Such result is expected for absorption in neutral media, resulting in observed single-scattered photons on bound electrons. The anticipation that the integrated simulation tables should converge in continuum to the integrated single-scattering events is confirmed by direct comparison with integrated Thomson single-scattering prescription. The Thomson scattering implementation in cone geometry was verified against Fig. 4b in \citet{Veledina2024}.
\\ \indent
For $i < \Theta$, the source is unobscured and primary radiation can be added to the reflected-only emission to provide an estimate of the total observed polarization in such geometries. We display the total polarization for the example of $\Theta = 70^\circ$, $i = 51^\circ$ and cone geometry in Fig. \ref{fig:energy_dep} for two different ionizations in order to illustrate that the unpolarized source dominates the emergent emission and we obtain $\lesssim 1\%$ of total polarization. For low ionization, the strong X-ray energy dependence of reflection spectra is more important than the energy dependence of the corresponding polarization, which imprints an increasing profile of the total polarization with energy across the X-ray band. Similarly, the superposition of the two components leads to energy independent and overall higher total polarization for more ionized reflectors than less ionized, contrary to the reflected-only polarization or total polarization for obscured sources (when $i > \Theta$). Unlike for partially transparent obscurers \citep[see the discussion in][]{Podgorny2024}, for fully opaque equatorial reflectors, the sign of the reflection-induced polarization does not alter with energy for any configuration. Thus, also the sign of the total polarization does not change with energy, if the source is unpolarized. 
\\
\\
In order to discuss the model parameter space efficiently, for the rest of the figures in this section, we display only the energy-integrated results between 3.5--6 keV, which avoids contamination by spectral lines. At the same time, we effectively examine the competing absorption and scattering effects within the 2--8 keV energy band of IXPE, the currently most sensitive operating X-ray polarimeter.
\\
\\
The choice of archetypal shape of the reflector affects the obtained polarization typically in the order of a few \%, slightly varying with energy, which is shown in the examples provided in Fig. \ref{fig:energy_dep}. In order to understand the impact of the curvature of the inner walls on the obtained polarization, we need to study contribution from different sections of the focusating reflectors, which is possible thanks to the imaging function in the {\tt torus\_integrator} routine. In Fig. \ref{fig:imaging} we show the observed images of the intermediately ionized examples from Fig. \ref{fig:energy_dep} (green lines) in the same geometrical configurations per panel. As for the explanation behind the general behavior of the integrated polarization on opening angle and inclination, which was provided in \citet{Podgorny2024} based on such imaging for the exemplary circular torus, nothing new appears for the elliptical torus, cone, and bowl. To illustrate this point, for the cone geometry only, we show the polarization degree, angle, and the polarized flux from each section of the observed reflecting surface, which can be compared to Fig. 6 of \citet{Podgorny2024}. In the rest of the panels of Fig. \ref{fig:imaging}, we show the difference in polarized flux and polarization degree between the studied shapes for each pixel, in order to explain the difference in the obtained total polarization via altering the curvature of the inner reflecting walls. The difference in polarization angles between the three geometrical archetypes are negligible ($\lesssim 1^\circ$) for all lines of sight.
\\ \indent
The difference between the torus and cone (central panels) and the bowl and cone (right panels) in polarized flux (blue and green shading) shows patterns fundamental to the shape definitions, which do not change with opening angle or global inclination. We see primarily the role of local inclination angles and proximity of each wall segment to the source in terms of illuminating, and thus reflected flux values, acting as weight factors to the spatially resolved polarization information. Close to the equator, the local incident inclination of the walls is the smallest for the torus and the largest for the bowl. Close to the upper edge of the reflecting area, it is the opposite. The walls of the bowl are generally further from the source than the walls of the torus (the cone being in between), but we view a projected picture. This means that even for intermediate inclinations, the difference in distances $r$ per pixel on the sky is smeared between different reflecting shapes, as we view, depending on the azimuth as well, generally more elevated sections of the torus at higher latitudes than of the bowl in identical lines of sight. The less inclined and closer sections to the source have higher ionization due to Eq. \ref{xidef} for the chosen constant-density ionization profile with $\beta = 2$. And the higher the ionization is, the more flux it reflects. Such areas are then contributing higher to the total emission, although their polarization is on average smaller (at least in the mid and hard X-rays, see Fig. \ref{fig:energy_dep}). The local emission inclination also plays a role due to the projection effects of reflecting surface areas and competing limb-brightening effects. However, its net impact on the reflected flux is secondary for intermediate inclinations compared to the impact of incident inclination angles. The local emission inclination angle relative to the surface normal generally depends on the azimuth, on the height above the equatorial plane, and on the adopted shape.
\\ \indent
To display the polarization degree part of the polarized flux in Fig. \ref{fig:imaging}, we display at each pixel a number and an additional color-coded arc, showing the polarization degree difference between the reflecting shapes. The polarization degree is, apart from local ionization, given by the local dominant scattering angle -- defined by the line of sight and the line between the center and the observed point on the surface. For scattering angles close to $90^\circ$, the locally emergent polarization fraction is the highest. Differences in local polarization between the torus and bowl occur on the back side of the structure from the observer, which mostly contributes to the reflected flux. The scattering angles in these regions are the largest for the torus and smallest for the bowl. However, for the combinations of $i$ and $\Theta$ shown, the scattering angles close to the upper rim already reach $90^\circ$, which means that the resulting polarization difference due to difference in scattering angles is small. At high latitudes, of higher order of importance for polarization are changing ionization effects. For the chosen ionization profile with $\beta = 2$, the local ionization drops by 1 order of magnitude from $\xi_0 = 500 \, \, \textrm{erg} \cdot \textrm{cm} \cdot \textrm{s}^{-1}$ at the equator to the upper illuminated edge. The relative ionization difference due to different shapes drives the polarization difference between the torus, cone, and bowl in the elevated regions close to the upper rim, which are dominating the observed radiation for small opening angles (ULX-like sources) where the rest of the inner illuminated walls of the funnel are obscured.
\\ \indent
For $\Theta = 70^\circ$ (bottom panels in Fig. \ref{fig:imaging}), it can be seen from the green and blue shading that such region only extends across the area where the flux contribution from the torus is lower than for the cone, and for the bowl higher than for the cone. Hence, for the torus the effect of enhanced polarization from the elevated regions above the equatorial plane is diminished. Closer to the equatorial plane, the torus shows lower polarization relative to cone (with higher relative flux contribution), just like the bowl relative to the cone (with lower relative flux contribution). Hence, there are some regions of the parameter space where the cone geometry does not always result in reflection-induced polarization fraction in between the torus and the bowl extremal shapes, which was depicted already in Fig. \ref{fig:energy_dep}. The reason behind is the dominant local ionization difference for the torus versus cone (closer distance and lower incident inclination of the torus walls with respect to the source), and the dominant local scattering angle difference for the bowl versus cone (lower scattering angle from the bowl walls given by the line of sight), competing with each other in the provided example. Although the depolarization for the torus versus cone at the equatorial ring disappears with azimuth, as the scattering angle differences become more important from the back to the left and right sides from observer's perspective, the disappearance with azimuth is not steep enough to prevent overall negative difference in reflection-induced polarization between the torus and cone for unobscured geometries. It was already noted in \citet{Podgorny2024} that the furthest side of the torus from the observer, which depolarizes the overall parallelly oriented polarization, is relatively narrow in the field of view compared to the left and right sides for high opening angles, which positively contribute to the total polarization.
\\
\\
In Fig. \ref{fig:xi_dep} we examine the role of $\xi_0$ and $\beta$ parameters. Higher overall ionization, defined primarily by $\xi_0$ at $\rho_\mathrm{in}$, reduces the observed reflected-only polarization (see Fig. \ref{fig:energy_dep}), which is due to the increased relative contribution of scattering to absorption in local reprocessing. The depolarization in 3.5--6 keV averaged values is gradual from nearly neutral reflection to nearly ionized reflection by about half the original values. The transition between neutral and ionized limits is shifted to lower/higher $\xi_0$ for lower/higher $\beta$, respectively, as $\beta$ defines the surface ionization profile further away from the equator via Eq. \ref{xidef}. For the total polarization, the changing relative flux contribution of the reflection component with ionization is again more significant in terms of the superposition of the reflected and the primary radiation than its changing polarization value. Thus, for an unpolarized source, the total (summed) polarization increases with increasing $\xi_0$ due to flux weighting of the two components.
\\ \indent
From the above analysis we discovered that ionization plays the largest role for small half-opening angles and large inclinations, when only a small stripe of the illuminated regions is visible. Indeed, because the small opening angles provide the largest difference in ionization between the equator and the upper rim of the illuminated surface, we obtain, e.g., depolarization from $\sim40\%$ to $\sim20\%$ for $\xi_0 = 1000 \, \, \textrm{erg} \cdot \textrm{cm} \cdot \textrm{s}^{-1}$, $\Theta = 40^\circ$ and $i = 79^\circ$ (brown dashed lines in Fig. \ref{fig:xi_dep}), when changing $\beta$ from 2 to -2. We note that for the smallest opening angles, the approximation of a single reprocessing event is the least valid, and we would expect multiple reflections inside the collimated funnels, generally depolarizing the emergent radiation due to less constrained geometry of scattering and smearing the differences between the tested shapes of the reflector. The difference in polarization due to change in shape between the cone, bowl and torus are generally dependent on $\xi_0$ and $\beta$, which can be seen in the central and right panels of Fig. \ref{fig:xi_dep}.
\\
\\
In Fig. \ref{fig:rho_dep}, we examine the role of remaining geometrical and source anisotropy parameters for a fixed ionization profile. We display the 3.5--6 keV polarization versus $\rho/\rho_\mathrm{in}$, which is effectively representing the skew of the inner walls. For large inclinations in obscured scenarios ($i > \Theta$), the skew is increasing polarization, as it enlarges the reflective area (the relative distance of the points at high latitudes with respect to the equatorial points) and creates a a higher ionization gradient from the equator to the upper rim. Lower ionization on the furthest points from the center pushes the summed polarization to higher values for either prevailing sign of the polarization. In addition, the skew affects reflection geometry and alters polarization even for a constant ionization profile in a non-trivial way, which was shown in Fig. 9 of \citet{Veledina2024}. For low inclinations in unobscured scenarios ($i < \Theta$) the changing skew has a significant impact on the reflection-induced polarization, which in the case of large opening angle (see the bottom left panel of Fig. \ref{fig:rho_dep} for the example of $\Theta = 70^\circ$ in cone geometry) changes the predominant orientation of polarization between $\rho/\rho_\mathrm{in} = 1$ and $\rho/\rho_\mathrm{in} = 2$. This is because the visible portions of the inner walls change their position from a cylinder to a wide-open cone, which alters the contribution from the sides versus from the back-center of the reflector, and changes the scattering angles. Nonetheless, the total polarization for such unobscured cases is still dominated by the source emission and is affected a little by changing $\rho/\rho_\mathrm{in}$.
\\ \indent
In Fig. \ref{fig:rho_dep}, we also show the impact of the bottom part of the reflecting structure by switching it on and off. Naturally, it has the largest effect when $\rho/\rho_\mathrm{in}\rightarrow1$, meaning that all studied geometrical shapes approach a pure cylinder. Then the skew of the observed part below the equator enables most of it to be observable. We refer to Appendix \ref{computations} for the discussion of the observability of such surface section for different $\Theta$ and $i$ per assumed geometry. Nonetheless, it can be estimated from Fig. \ref{fig:rho_dep} that the reflection from the half-space below the equator of the optically thick structure has a negligible impact on the total polarization, unless we are in very specific configurations.
\\ \indent
When comparing the three different examples of source emission in Fig. \ref{fig:rho_dep}, there are two effects in play: changing the intrinsic anisotropy of the flux and polarization. The cases of isotropic and cosine source are both unpolarized, but the cosine source is emitting with less flux towards the equator and more towards the poles. Comparing the effect of the two for the cone in Fig. \ref{fig:rho_dep}, we see that for high inclination $i = 79^\circ$ (dashed lines), the difference is smaller for $\Theta = 40^\circ$ than for $\Theta = 70^\circ$ when the reflector approaches rather a ring geometry close to the equatorial plane. In such case, the nearly equatorial regions with the highest illumination difference due to source flux anisotropy can still be observable. For the cosine source the flux is less towards the equator than for the isotropic case, thus the ionization of such regions is lower, hence the reflection-induced polarization is higher, which can be seen in the bottom left panel of Fig. \ref{fig:rho_dep}. While for the displayed case of $\Theta = 40^\circ$, the equatorial section is self-obscured and we observe the elevated parts of the structure illuminated by angles smaller than $60^\circ$ from the axis, which are more ionized for the cosine source compared to the isotropic source, thus inducing lower polarization. For low inclination $i = 28^\circ$ (dotted lines), we observe the equatorial regions even for $\Theta = 40^\circ$, which reverses the effect of cosine-source anisotropy on reflected-only polarization. The equatorial sections are more exposed for higher $\rho/\rho_\mathrm{in}$, which creates a larger polarization difference (with a sign change even) in the purple and orange dotted lines in the top left panel of Fig. \ref{fig:rho_dep} for $\rho/\rho_\mathrm{in} \sim 3$ than for $\rho/\rho_\mathrm{in} = 1$. At higher $\rho/\rho_\mathrm{in}$, the polarization from the cosine source converges again to the isotropic case.
\\ \indent
If the source is additionally polarized, it will imprint on the reflected polarization. We refer to Appendix D of \cite{Podgorny2024} for a detailed discussion of this effect. In here we test a case of a semi-realistic polarized source in the slab-corona prescription in blue lines in Fig. \ref{fig:rho_dep}. The prescription (\ref{slab_prescription}) encodes anisotropy in both flux and polarization. The flux drops with higher emission angle from the axis, qualitatively like the cosine source. The polarization is aligned with the axis and increases from 0\% towards the poles to $\sim 9\%$ at $80^\circ$ emission angle and then decreases again to 0\%, which is not important, as the equatorial regions are effectively unilluminated. Compared to the isotropic and cosine source, the positive intrinsic polarization adds reflection-induced polarization in the net positive 3.5--6 keV values and subtracts polarization in the net negative 3.5--6 keV values, nearly independently of $\rho/\rho_\mathrm{in}$ and by a few \%, depending on geometry, which is approximately the intrinsic angle-averaged polarization fraction. The only exception in Fig. \ref{fig:rho_dep} is the low inclination and low half-opening angle case, where the slab-corona source acts similarly to the cosine source and is more polarized compared to the isotropic source, despite the net negative polarization. This is because in such scenarios the flux anisotropy plays a larger role than the intrinsic polarization effects and we obtain higher flux contribution from the equatorial, more ionized and less polarizing regions in the isotropic case. Out of the blue and orange dotted lines in the top left panel of Fig. \ref{fig:rho_dep} with similar flux anisotropy laws, the slab-corona induces lower polarization, as its intrinsic polarization effectively subtracts from the integrated result, compared to the unpolarized cosine source. In such unobscured geometries, the observed total emission is anyway dominated by the primary component. This is also the reason why the slab corona is the only source type out of the three studied, which promises for unobscured accreting compact objects detectable 2--8 keV polarization with contemporary X-ray polarimeters. Its total polarization for $i = 28^\circ$ (solid lines) significantly deviates from the unpolarized sources in the left panels of Fig. \ref{fig:rho_dep} due to intrinsic polarization of the source.
\\ \indent
The central and right columns of Fig. \ref{fig:rho_dep} show again the difference in the integrated polarization degree between the three studied shapes of the reflector in the geometrical and anisotropy parameters of the model. Even for a constant ionization across the reflector, the anisotropy changes the weighting of different reflecting sections of the inner walls. That is why the purple dotted lines deviate strongly for the torus, where the equatorial regions show negative polarization difference compared to the cone (cf. Fig. \ref{fig:imaging}), because the isotropic source enhances the equatorial contributions compared to the slab corona and the cosine source. The skew of the inner walls is strongly related to the impact of the curvature of the walls on reflection-induced polarization. We see that regardless of the inclination and opening angle, the differences in observed polarization converge to 0\% for $\rho/\rho_\mathrm{in} \rightarrow 1$, where the cone, the torus, and the bowl, all effectively become a cylinder. 
\\ \indent
In order to visualize the relative importance of curvature of the inner walls in the full parameter space in the two most important geometrical parameters, the inclination and the half-opening angle, we provide the reflected-only 3.5--6 keV polarization in Fig. \ref{fig:heatmaps} for two values of $\rho/\rho_\mathrm{in}$ and all allowed combinations of $i$ and $\Theta$ that are not causing a full eclipse. On the left panels for absolute values of polarization in the cone geometry, we see a well-confirmed pattern of reflection-induced polarization in the $\{i,\Theta\}$ space \citep{Ursini2023, Podgorny2024, Veledina2024, Meulen2024}. We newly test ionization effects and confirm that the shift of the 2D peak (in its shape resembling a heron's head) moves towards the grazing-angle line for higher $\rho/\rho_\mathrm{in}$, as greater skew enables lower ionization of the parts at higher latitudes, increasing the reflection-induced polarization. Similarly, the reflection features in polarization for unobscured geometries shift closer to the grazing-angle line for higher $\rho/\rho_\mathrm{in}$, which is the region where scattering from the bottom part of the structure (at $z < 0$) is affecting the results (cf. Fig. \ref{fig:torus_curves}). The relative differences in observed polarization fraction between the torus-like and bowl-like focusation scenarios can be as large as 30\% in some regions of the parameter space. The self-obscuration by the front elevated side of the torus (dashed purple line in Fig. \ref{fig:geometry_sketch}) causes shadowing of the reflecting sections closer to the equator, resulting in a strong negative polarization difference for high inclinations and high $\rho/\rho_\mathrm{in}$ (white region in the bottom panel for the torus versus cone in Fig. \ref{fig:heatmaps}) from the cone, where such self-obscuration does not apply. This is the anticipated direction of change of the results, should we consider further elevated optically thick accretion material in the bowl and cone geometries, which may be convex after an inflection point at some height \citep[forming a large-scale cusp, see geometry indications in, e.g.,][considering the full provided range in radius]{Abramowicz1978, xrism2024}. The further-most regions of puffed matter should physically adhere back to the equatorial plane with loss of gravitational impact of the central object, and despite not being directly illuminated by a central source of X-rays, they might intervene in obscuration related changes in the observed polarization fraction.

\section{Relaxing the optically thick assumption}
\label{sec:discussion}

\begin{figure} 
\centering
\includegraphics[width=0.4\linewidth, trim={3.8cm 5.5cm 17.5cm 5.2cm}, clip]
{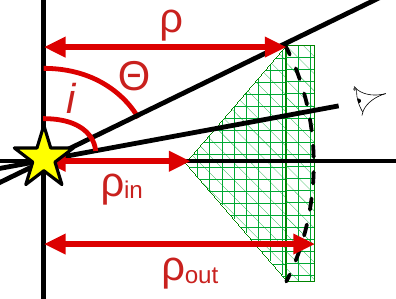}
\caption{Sketch of the geometry for the studied example of the partially transparent homogeneous equatorial medium in cone-like geometry. The only new parameter with respect to the optically thick model is $\rho_\mathrm{out}$, which we set to equal the distance between the center and the tangential points between the obscuring structure and the cone dissecting its half-opening angle. Otherwise, we operate in an identical setup and notation to the optically thick scenario.} \label{fig:thin_cone_sketch}
\end{figure}

For the winds and dusty tori near accreting compact objects, we often anticipate and spectroscopically detect partial X-ray transparency of the reflecting (and absorbing) equatorial medium. Having provided an in-depth examination of optically thick reflection in terms of shape and ionization in the previous section, here we discuss a so-far poorly articulated link to the studies examining optically thin reprocessing in more detail. We derive a continuous range of examples with varying optical depth from the fully opaque model examples presented above, to the optically thin scenarios, recently explored in X-ray polarization via modelling by various authors. The work of \citet{Ratheesh2021} provided examples of a fully ionized and fully neutral scenario, but only specific cases for equatorial column density were shown. The investigation was enlarged in the studies \citet{Podgorny2023b, Podgorny2024}, which were later on independently confirmed in related conclusions by a narrower, but more in-depth study of \citet{Meulen2024}. Regarding the discussion of partial ionization for transparent obscurers in \citet{Podgorny2023b, Podgorny2024}, only a mixture of neutral absorbers with free electrons was used as an approximation to a more realistic ionization structure. Regarding partial transparency, only specific cases of equatorial column density were shown in \citet{Podgorny2023b, Podgorny2024} to illustrate the discussion. \citet{Meulen2024} provided a finer grid of computations with respect to optical depth, but only for a cold medium. Polarization from centrally illuminated optically thin clouds was analytically studied by, e.g., \citet{Gnedin1973,Dolginov1974, Brown1977}, and generalized by, e.g., \citet{Brown1978,Simmons1982,Cassinelli1987,Brown1989}. The studies of \citet{Tomaru2024, Nitindala2025}, departing from classical results of \citet{SunyaevTitarchuk1985}, provided specific exploration of centrally illuminated electron-scattering dominated equatorial accretion-disc winds with varying optical depth, but without the competing energy-dependent photo-electric absorption effects. 

In here, we make a clear computational and visualization link from the fully ionized and fully neutral optically thick scenarios, presented with all intermediate (and consistently computed) ionization cases in the previous section, to the optically thin solutions, via additional, pure Monte Carlo simulations in a specific cone geometry and only in the strict fully ionized and fully neutral limits. Such analysis illustrates in the two extreme ionization examples and one obscuration geometry ($i > \Theta$) what to anticipate from relaxing the optically thick assumption held in the previous section. It is out of scope of this work to discuss X-ray polarization for the complex combinations of geometry and partial ionization effects with respect to partial transparency in detail to repeat the conclusions from the abovementioned studies, nor we provide a yet missing X-ray polarization prediction for a self-consistent 3D ionization structure solved via non-LTE radiative transfer for a partially transparent obscurer. But we expect that the discussion with respect to energy-dependence of the resulting X-ray polarization is rather universal and applicable for other geometries and partial ionization 3D profiles, which was to a limited extent explored in \citet{Podgorny2023b, Tomaru2024, Podgorny2024}. We note that similarly to the thick obscurers, optically thin winds, without any further constraints on the source anisotropy, the wind opening-angle, optical depth distribution, and inclination, can produce a large diversity of observable polarization fractions with either prevailing sign \citep{Nitindala2025}. Even when considering both observational and theoretical arguments on the expected wind structure for various accretion conditions, adding single-scattering interaction with large-scale winds to the primary emission is a flexible way to explain various energy-integrated IXPE measurements of accreting compact objects \citep{Nitindala2025}. Comparing the expected energy-dependence of X-ray polarization from modelling partially transparent obscurers with energy-resolved X-ray polarimetric data can be a constraint for such efforts, should the energy-dependence of polarization arise directly due to interaction with the wind structure and not due to X-ray emitting and reprocessing components (and their combinations), which are located closer to the compact object.

\begin{figure}[ht!]
\centering
\begin{tikzpicture}[
x=1pt, y=1pt,
inner sep=0pt,
outer sep=0pt]
\node[anchor=south west] (base) at (0,0)
  {\includegraphics[width=1\linewidth, trim={0.1cm 0.2cm 0.2cm 0.2cm}, clip]
{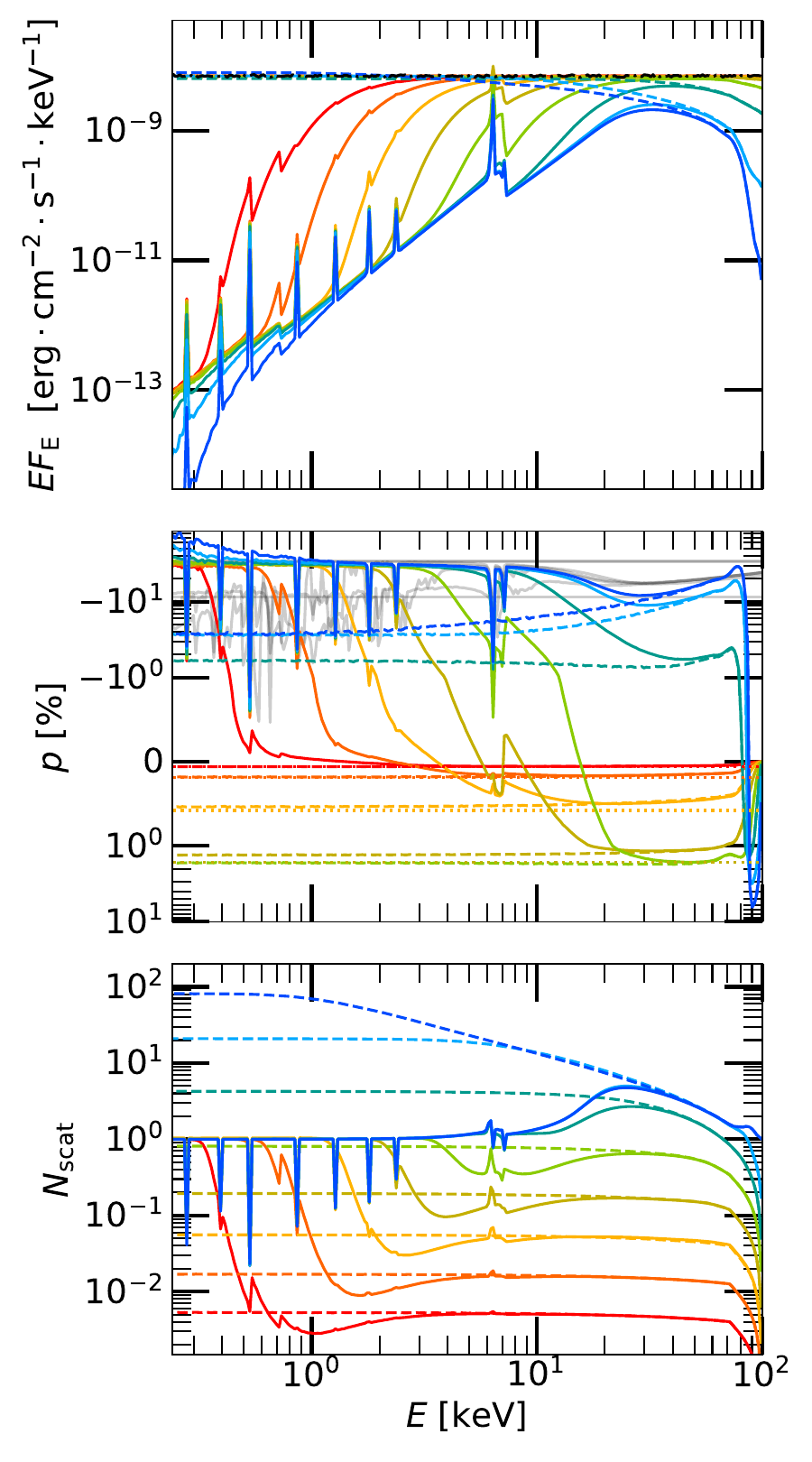}};
\node[anchor=south west] at (2,175)
  {\includegraphics[width=0.06\linewidth]{para.png}};
\node[anchor=south west] at (2,269)
  {\includegraphics[width=0.08\linewidth]{perp.png}};
\end{tikzpicture}
\caption{The results of the additional Monte Carlo simulation for $i = 50^\circ$ and the partially transparent cone with $\Theta = 40^\circ$ and $\rho = \rho_\mathrm{c}$, illuminated by a central unpolarized isotropic primary power-law radiation with $\Gamma = 2$. We show the observed total spectra (top), $EF_\mathrm{E}$, corrected for the slope of primary radiation, the corresponding total polarization degree (middle), $p$, and the average number of scattering events before detection (bottom), $N_\mathrm{scat}$, versus energy. In the color code, we show from red to blue different cases of equatorial column density: $N_\mathrm{H} = 10^{22}, 10^{22.5}, 10^{23}, 10^{23.5}, 10^{24}, 10^{24.5}, 10^{25}, 10^{25.5}\, \textrm{cm}^{-2}$, respectively. The solid lines represent a fully neutral obscurer, the dashed lines represent a fully ionized obscurer. In dotted lines we show examples of the energy-independent analytical polarization prediction from \citet{Brown1977} valid for low $N_\mathrm{H}$ and full ionization. In dot-and-dash black lines, we show the intrinsic emission. In light gray in the background, we overplot all polarization curves from the middle left panel in Fig. \ref{fig:energy_dep}. All spectra are renormalized to intrinsic luminosity $L_\mathrm{X} = 10^{37} \, \textrm{erg}/\textrm{s}$ and distance $D = 3 \, \textrm{kpc}$.} \label{fig:NH_effects}
\end{figure}

Fig. \ref{fig:thin_cone_sketch} shows the adopted geometry. We use a power-law point-like source with unpolarized isotropic emission. Only one extra parameter to the cone geometry, defined for optically thick reflection in Sec. \ref{sec:model}, is used to effectively investigate the role of optical depth, which is the outer radius $\rho_\mathrm{out}$. We assume a homogeneous axially symmetric medium, mirror symmetric to both sides of the equator, and filled either with neutral atoms with bound electrons, according to solar abundances from \cite{Asplund2005} with $A_\mathrm{Fe} = 1.0$, or with free electrons only, assuming their density according to the same atomic abundances, as if all species were fully ionized. The equatorial column density is then defined as $N_\mathrm{H} = n_\mathrm{H} \cdot (\rho_\mathrm{out}-\rho_\mathrm{in})$, where $n_\mathrm{H}$ is the hydrogen number density. We use the {\tt STOKES} code, version 2.35, which enables the geometrically thick cone geometry. Similar computations, using the same method, were already used to support the observational interpretations in \citet{Veledina2024,Veledina2024b} regarding partially transparent accretion funnels, but they were never published in a quantitative and sufficiently explanatory extent.
\\
\\
In Fig. \ref{fig:NH_effects}, we show the obtained spectra, total polarization\footnote{Note that due to the necessity of displaying all obtained dependences clearly, we use the symmetric logarithm scale for polarization in the middle panel of Fig. \ref{fig:NH_effects}, which is a logarithm scale applied to each sign of polarization with a linear scale between -1\% and 1\%.}, and average number of scatterings versus energy for changing $N_\mathrm{H}$. The system is viewed in the exact same geometry and source properties as in Fig. \ref{fig:energy_dep} for the cone with $\Theta = 40^\circ$. Therefore, we provide in the background also the polarization results of the optically thick model in all intermediate and self-consistently computed cases of partial ionization, presented in the middle left panel of Fig. \ref{fig:energy_dep}, to which the partially transparent results converge for high $N_\mathrm{H}$. The remaining discrepancies between the highest $N_\mathrm{H}$ case provided and the optically thick model are due to remaining partial transparency in the most shallow and elevated layers of the adopted conical obscurer and due to multiple reflections within the inner walls of the cone, which were not taken into account for the optically thick model presented in the previous section and which cause a depolarization for high ionization.

The optically thin fully ionized results are verified in polarization predictions against the analytical approximations provided in \citet{Gnedin1973,Dolginov1974,Brown1977} (dotted lines in Fig. \ref{fig:NH_effects}), which are valid for low $N_\mathrm{H}$.\footnote{We adopted the general single-scattering formulae (17)--(20) from \citet{Brown1977}. Note that the simplified Eq. (23) presented as a main result in \citet{Brown1977} omits a factor of 2 due to an algebraic error. The same result (up to the factor of 2) was independently derived already in \citet[][Section 2]{Gnedin1973}, and re-formulated in \citet[][Eq. (14)]{Dolginov1974}.} A point-like isotropic central source of emission is assumed. Energy-independent Thomson scattering in an axially symmetric cloud of electrons is taken into account. We calculated the values for the same electron density distribution as in our Monte Carlo simulation.\footnote{We could have also assumed only one half of the mirror-symmetric scattering region, above or below the equatorial plane where, e.g., a geometrically thin and optically thick accretion disc would be located. Then the analytical formulae show that due to the mirror symmetry, the resulting polarization degree is smaller by a factor of 2, if the entire optically thin sub-region below or above the equatorial plane is omitted.} For $N_\mathrm{H} \rightarrow 0$, both the fully ionized and the fully neutral cases converge to the naked source emission in spectra and polarization, which is shown in dot-and-dash black lines in Fig. \ref{fig:NH_effects}.

We refrain from discussing the expected sign and magnitude of polarization for particular accretion scenarios, as its extremal values with respect to $N_\mathrm{H}$ are dependent on the assumed geometry, density profile, and source setup. Fig. \ref{fig:NH_effects} shows only one example to illustrate the effect of relaxing the optically thick assumption in the fully ionized and fully neutral limits, particularly focusing on the related X-ray polarization energy-dependence, which can be well tested by instruments such as those on board of IXPE and its successors. For the partially transparent results (in color in Fig. \ref{fig:NH_effects}), a more detailed treatment of the ionization structure and kinematics than allowed by our simplistic model presented in this section would result in synthetic absorption lines or P-Cygni line profiles, often observed in partially ionized winds. The example calculated in \citet{Tomaru2024} for a nearly fully ionized wind, however, suggests a little impact of the related polarization absorption-line change on the observable energy-resolved polarization with IXPE capabilities and Galactic X-ray sources.

The fully ionized results agree with energy-integrated polarization presented in Fig. 9 of \citet{Podgorny2024}, although here we assume a different reflector's shape and take into account zero-scattered photons, i.e. the primary radiation, which depolarizes the reflected contribution for low $N_\mathrm{H}$. This results in a peak positive total polarization for $N_\mathrm{H} \sim 10^{24} \, \textrm{cm}^{-2}$, and for higher $N_\mathrm{H}$ multiple scatterings begin to depolarize the result. The prevailing alignment of the polarization angle with the system axis of symmetry for low $N_\mathrm{H}$ is due to majority of the scatterings occurring close to the equatorial plane when the system is effectively optically thin, which for any intermediately inclined observer and isotropic source results in average scattering plane aligned with the equatorial plane \citep[e.g.,][]{SunyaevTitarchuk1985, Podgorny2023b, Podgorny2024}. The same orientation of polarization for optically thin fully ionized winds was obtained in independent studies of \cite{Tomaru2024, Nitindala2025}, but a detailed quantitative comparison\footnote{We note that the two studies are not in mutual agreement on the general sign and magnitude of X-ray polarization expected via wind scattering effects for electron-scattering disc sources, likely due to different density structures assumed, which results in different conclusions presented therein about interpretation of particular IXPE data.} is out of the scope of this work. For a strong anisotropy of the source, increased relative contribution of scattering in elevated regions of an optically thin wind at high latitudes, may also result in negative observed polarization \citep{Nitindala2025}. The allowed passage of photons through the obscurer is also the reason why for optically thin obscurers the impact of the bottom half-space ($z < 0$) on the observed X-ray polarization is expected to be more significant than for the optically thick obscuration, where such regions are contributing only from a limited range of inclinations (see Fig. \ref{fig:torus_curves}).

For extreme $N_\mathrm{H}$ and full ionization, the photons are trapped inside the thick electron-scattering cloud and are more likely to escape from the inner side where they entered the scattering medium. Hence, the fully ionized curves converge to the optically thick reflecting cone model, resulting in negative polarization for this combination of geometry, inclination, and source properties. The dense electron clouds do not result in energy-independent polarization. The reason is that inside the simulation, we use Compton down-scattering instead of elastic Thomson scattering, which reduces the photon's energy upon each scattering event. Then towards the hard X-rays, we observe incrementally fewer photons with respect to the spectral energy distribution of the source. Those high-energy photons that remain are polarized on average higher than elastic scattering predicts, because they undergo less number of scatterings, which is observed from the top and bottom panels of blue curves in Fig. \ref{fig:NH_effects}. The increase in obtained polarization is rather steady, but we note that the results would be different if a counter effect of Compton up-scattering in hot plasma was included, or if we included special relativistic effects due to more realistic plasma kinematics, or if a different high-energy cut-off was used.

In the fully neutral scenario, the energy dependence of the resulting polarization is strong due to the strong energy dependence of absorption opacity, which can be indirectly traced in the shown spectra and number of scatterings. At low energies and/or high $N_\mathrm{H}$, the average number of scatterings for neutral obscurers converges to 1, as photons are either absorbed, or can scatter once from the bound electrons back into the half-space with vacuum, dissected by a sharp density transition. In other words, the single-scattering approximation for integration across the inner walls of the reflector is well justified for low energies and/or high $N_\mathrm{H}$. The related energy-dependent transition in predominant polarization orientation from optically thick and negatively polarized results at soft X-rays to optically thin and positively polarized results at hard X-rays has been extensively studied for neutral or nearly neutral obscurers and isotropic sources, for example, in \citet{Marin2018b, Marin2018c, Podgorny2023b, Podgorny2024b, Podgorny2024, Meulen2024}. We emphasize here the steepness of the polarization energy transition from the nearly energy-independent (apart from spectral lines) opaque soft X-rays to nearly energy-independent (apart from down-scattering effects) transparent hard X-rays, where the fully ionized and fully neutral results converge in all three quantities shown in Fig. \ref{fig:NH_effects}, as the passing photons effectively register only the scattering cross-section. The steepness in the mid X-rays will, to a large extent, remain for partially ionized intermediate cases, as increasing ionization first affects reflection or transmission spectra in the soft X-rays.
\\
\\
Our results indicate two possibilities for the emergence of the small and approximately linear increase of continuum polarization between 2 and 8 keV seen in many IXPE observations of accreting compact objects \citep{Dovciak2024, Marin2024, Poutanen2024, Ursini2024}, due to the interaction of isotropically emitted X-rays with homogeneous equatorial winds only. One mechanism is Compton down-scattering, but we observe noticeable effects in the 2--8 keV only for optically thick electron-scattering clouds with high multiple-scattering contributions. The other mechanism is the right energy-dependent balance between absorption and scattering for neutral or partially ionized wind, causing a decrease of continuum polarization with energy for negatively polarized part of the spectrum (the optically thick regime at soft energies for geometrically thick obscurers or low inclined observers) and increase of polarization for positively polarized part of the spectrum (the optically thin regime at hard energies for any geometry or any regime for high opening angles and high inclinations). We refer to \citet{Podgorny2024} (Fig. 3) for more details. Although the profile of integrated polarization with energy is highly sensitive to the ratio of scattering and absorption opacities, given in general by the spatial distribution of the gas composition and the source properties, a small increase in polarization degree with energy can be expected from the interaction with nearly fully ionized winds in certain configurations.

For unpolarized isotropic sources and homogeneous obscurers, the polarization energy transition between the optically thick and optically thin regimes, if applicable, would have to be below 2 keV, with the inclination and wind geometry parameters matching both polarization and spectroscopic properties of each source. In the configurations with the transition below 2 keV or with positive polarization across the X-ray band, the polarization may slightly increase \citep[see Fig. \ref{fig:NH_effects} or e.g.,][]{Marin2018b, Marin2018c, Podgorny2024b, Podgorny2024, Meulen2024}, although the slope of polarization degree with energy in 2--8 keV still depends on the assumed obscurer's composition, density, geometry, and the inclination. 

For anisotropic cosine sources and optically thin winds, as shown in \citet{Nitindala2025}, the regime of negative total observed polarization may apply. In such case, a non-homogeneous distribution of absorbers could help to restore the increasing polarization fraction with energy. Increased absorption in the elevated parts of the wind at high latitudes will allow higher relative contribution from the scatterings closer to the equatorial plane, which can result in depolarization at soft energies. The same effect will be achieved by energy-dependent emission properties of the source. It was proposed in \citet{Nitindala2025} that relativistic effects at the source, causing energy-dependent boosting towards the equator, could result in increasing X-ray polarization fraction with energy after taking into account the interaction with the large-scale winds, while keeping the negative polarization sign. According to our qualitative estimates, such luminosity anisotropy would result rather in depolarization of net negative polarization at higher energies, unless such flux anisotropy is compensated and overcome by increased intrinsic negative polarization towards the equator, which is generally expected from electron-scattering discs \citep{Chandrasekhar1960}. However, the exact required X-ray source properties can neither be supported nor disfavored by the computations presented in this work.

The modelling discussion could be significantly constrained by both, more realistic source-wind models, as well as future X-ray polarimetric data for accreting compact objects outside the IXPE band, such as from the XL-Calibur experiment operating in 15--80 keV \citep{Awaki2025}. Monte Carlo simulations are a useful tool for examining the effects of multiple scatterings in density structures obtained from hydrodynamical simulations \citep{Tanimoto2023,Tomaru2024} and the X-ray observations suggesting complex clumpy accretion-disc wind structures \citep{xrism2025,xiang2025}. The {\tt STOKES} code employed in this study supports coupling with \text{non-LTE} radiative transfer iterative solvers, establishing it as one of the leading X-ray polarization Monte Carlo tools currently available for modeling complex ionized 3D environments.

\section{Conclusions}
\label{sec:summary}

We built an X-ray spectro-polarimetric model capable of examining the role of partial ionization and the shape of equatorial optically thick reflecting media around accreting compact objects. By theoretically examining different geometrical and ionization configurations, we conclude that both affect the observed reflection-induced polarization. Albeit in many configurations, it is difficult to disentangle the impact of changing local scattering angles, limb-brightening, or ionization, it is optimistic for the current and forthcoming X-ray polarimeters that the designed parameter space leads to diverse polarization properties of compact objects enclosed by geometrically and optically thick material, resulting up to tens of \% of X-ray polarization in both parallel and perpendicular polarization with the axis. We tested that the source properties, such as anisotropy and intrinsic polarization, are also imprinted in the reflected emission. In order to fully examine the detectability of the introduced features, a construction of a tabulated {\tt XSPEC} model is required. Such ambitions, as well as detailed model-data comparisons and data fitting with the updated {\tt xsstokes\_torus} model, using the IXPE archival observations, are left for follow-up studies.

One of the parameters of our simplified model was the curvature of the inner walls of the reflector, studied in three extremes: an elliptical torus, a cone, and a bowl. Although in some, if not most, regions of the parameter space, the adopted shape of the reflector plays a marginal role, in others it results in observed 2--8 keV polarization difference by more than 30\%, significantly deviating the inferred estimates on opening angle and inclination from X-ray polarization reflection signatures. Highly sensitive to such geometry assumptions are configurations with small half-opening angles (such as ULXs) viewed obscured and under small inclinations, which is e.g. the configuration suggested for the Cygnus X-3 system in \citet{Veledina2024, Veledina2024b}, which may bring additional uncertainties to the polarimetrically derived opening angle. However, this is the region of the parameter space where the neglected multiple reflections inside the collimating funnel will mostly apply, which reduces the decisiveness of the presented calculations. 

Further limitations of our model include a strict compactness of the source, located in the center of the system, and the neglect of relativistic effects. The latter is especially important in the inner-accreting regions, where our results should not be taken as a definitive observational prediction. Instead, we shed light on the impact of the shape and ionization of the accreted material, which are only two of the many ingredients in the calculation of emission from complex structures occurring near compact objects. The presented study can thus serve as a base reference for more sophisticated simulations.

Lastly, we discussed the impact of relaxing the optically thick assumption for the obscuring medium, suitable for examination of e.g. accretion-disc winds or Compton-thin dusty tori. Although an exact polarization profile with energy for a partially ionized and transparent wind requires a self-consistent ionization structure pre-computation, the results for a detailed ionization treatment in the optically thick regime (Sec. \ref{sec:results}), the results in fully neutral and fully ionized limits with varying optical depth (Sec. \ref{sec:discussion}), and the results for mixed neutral and fully ionized two-phase partially transparent media \citep{Podgorny2023b, Podgorny2024} put into challenge the interpretations of the IXPE data that the ubiquitously observed increase of polarization degree with energy is solely due to scattering (and absorption) in large-scale equatorial winds. An energy-dependent anisotropy of the source in combination with scattering inside a wind, or inhomogeneity of the ionization structure of the wind are likely a condition for the models to align with some of the IXPE observations, where total perpendicular polarization angle to the projected system axis is assumed, with rather specific than generic configurations to be required.

\begin{acknowledgements}

J.P. thanks Michal Dovčiak, Frédéric Marin, René Goosmann, and Petr Kurfürst for occasional, but extremely useful discussions during the progress of this work, and acknowledges institutional support from RVO:67985815 and the usage of computational facilities at the Silesian University in Opava. 

\end{acknowledgements}

\bibliography{refs}
\bibliographystyle{yahapj}

\begin{appendix}
\section{Geometrical implementation}\label{computations}

In this section we provide details on the numerical integration of the local reflection tables introduced in \cite{Podgorny2022}, including their neutral version introduced in this work, for the various adopted geometries. For each case, we use the original energy binning, power-law index, and primary polarization state of the tables. For the case of slab-corona illumination, the local reflection tables computed for 3 independent polarization states of the primary are linearly transformed via Eq. A2 from \cite{Podgorny2024} at each point of the surface to account for the prescribed polarization of incident emission.

\subsection{Elliptical and circular torus}

We first introduce the elliptical torus, which reduces to the circular case, described in \cite{Podgorny2024}, if the semi-axes $a$ and $b$ of the ellipse in the meridional plane are equal. In the Cartesian coordinate system $\{x,y,z\}$ with base vectors $\Vec{e}_\textrm{x} = (1,0,0)$, $\Vec{e}_\textrm{y} = (0,1,0)$, $\Vec{e}_\textrm{z} = (0,0,1)$ we use the standard elliptical torus surface parametrization
\begin{equation}\label{coordtrafo}
	\begin{aligned}
		x &= (R+a\cos{v})\cos{u} \\
		y &= (R+a\cos{v})\sin{u} \\
            z &= b\sin{v} \textrm{ ,}
	\end{aligned}
\end{equation}
where $R = a + \rho_\textrm{in}$ is the distance between (0,0,0) and the center of the toroidal ellipse. These are the transformation relations to \{$u$, $v$\} coordinates with base vectors $\frac{\partial}{\partial u} = (-\sin{u}, \cos{u}, 0)$, $\frac{\partial}{\partial v} = (-\sin{v}\cos{u}, -\sin{v}\sin{u}, \cos{v})$. For a given $\Theta$ and $\rho$,
\begin{equation}
   \begin{aligned}
    a &= \frac{\rho - \rho_\mathrm{in}}{2\rho_\mathrm{in}-\rho} \, \rho_\mathrm{in} \,\, , \\
    b &= \sqrt{\frac{\rho}{a+\rho_\mathrm{in} - \rho}} \,a\,\cot{\Theta} \, \, .
\end{aligned}
\end{equation}
The implicit formula for the surface is
\begin{equation}
\Phi = \left(x^2 + y^2 + z^2\, \frac{a^2}{b^2} + R^2 - a^2\right)^2 - 4R^2(x^2+y^2) = 0 \textrm{ .}
\end{equation}

The emergent Stokes parameters $I$, $Q$, and $U$, tabulated for local reflection, are linearly interpolated in local reflection angles $\mu_\textrm{i} = \cos{\delta_\textrm{i}}$, $\mu_\textrm{e} = \cos{\delta_\textrm{e}}$, and $\Phi_\textrm{e}$ \citep[for a sketch see Fig. B1, left panel, in][]{Podgorny2024} measured in the local frame with respect to the normal $\Vec{n} = \frac{\nabla \Phi}{\lvert \nabla \Phi \rvert}$, the incident vector $\Vec{I} = (x, y, z)$, the emission vector $\Vec{E} = (0, \sin i, \cos i)$, and their projections to the tangent plane of the surface at each \{$u$, $v$\}. The emergent polarization vector is for each point additionally rotated to conform to the global definition of polarization angle with respect to the projected system axis to the polarization plane perpendicular to the reflected photon's momentum. We do not account for those points, which are not directly illuminated by the source, i.e. where $\delta_\textrm{i} \notin [0; \pi/2]$. This condition for the $v$ coordinate results in a reduced integration range $\pi - \arcsin{(\rho\cot{\Theta}/b)} < v < \pi + \arcsin{(-\rho \cot{\Theta}/b})$, for $B = 1$, or $\pi - \arcsin{(\rho\cot{\Theta}/b)} < v < \pi$, for $B = 0$. The integration range in $u$ coordinate is $\pi/2 < u \leq 3\pi/2$, while the other half-space $3\pi/2 < u \leq 5\pi/2$ is added symmetrically with the opposite sign of the resulting Stokes parameter $U$. In the reduced ranges, we opted for $N_\mathrm{v} = 80$ and $N_\mathrm{u} = 50$ linearly spread points, which produces converging results to an exact solution for a majority of the parameter space explored. In extreme cases of small $\Theta$ and large $i$, we had to use a finer integration grid.

For the numerical integration, each reflecting point $\{u, v\}$ needs to be weighted by the area corresponding to the edges of the bins $u_1, u_2, v_1$, and $v_2$ in the aforementioned ranges and linear binning. We approximate the contributing local reflecting surface by the area of a tangent rectangle to the elliptical torus:
\begin{equation}
\begin{split}
    A_\textrm{u,v} & = \int_{u_1}^{u_2} \int_{v_1}^{v_2} (R+a\cos v) \sqrt{a^2 \sin^2v + b^2\cos^2v} \, \mathrm{d}v \, \mathrm{d}u \\
    & \approx (R+a\cos v) \sqrt{a^2 \sin^2v + b^2\cos^2v}\, (v_2-v_1)(u_2-u_1) \textrm{ .}
\end{split}
\end{equation}

The illuminated segments need to be visible by the observer, i.e. $\delta_\textrm{e} \in [0; \pi/2]$. This further reduces the integration range in $v$ due to shadowing up to
\begin{equation}\label{vlimit}
    v_\textrm{limit}(u,i) = -\arctan \left( \frac{b}{a} \sin u \tan i \right) + \pi \textrm{ ,}
\end{equation}
which is derived from the opacity condition $0 = \Vec{n} \cdot \Vec{E}$. Note that in the equivalent Eq. B4 in \cite{Podgorny2024} for a circular torus was an erroneous factor 2, which however has a negligible impact on the results presented therein. 

In addition, to avoid those points, which are self-obscured by a closer side of the torus towards the observer, we have to solve a set of parametric equations $\Vec{X}_1 + t\Vec{E} = \Vec{X}_2$, where $\Vec{X}_1=$ ($x_1$, $y_1$, $z_1$) begins at the further side from the observer and is given by $u$ and $v_\textrm{self-obs}(u,i)$, which is the unknown limit for the integration range in $v$ due to self-obscuration. The emission at $\Vec{X}_1$ is directed towards the observer in the $\Vec{E}$ direction, which is tangential to the torus surface at the observer's side at a point $\Vec{X}_2=$ ($x_2$, $y_2$, $z_2$) defined by $v_\textrm{limit}(u_\mathrm{t},i)$ for unknown $u_\textrm{t}$. The set of equations for an elliptical torus parametrized as (\ref{coordtrafo}) yields
\begin{equation}\label{cond2}
	\begin{aligned}
		0 &= (R + a\cos v_\textrm{self-obs})\cos u - (R + a\cos v_\textrm{limit}(u_\textrm{t},i) )\cos u_\textrm{t} \\
		0 &= (R + a\cos v_\textrm{self-obs})\sin u + t\sin i - (R + a\cos v_\textrm{limit}(u_\textrm{t},i) )\sin u_\textrm{t} \\
            0 &= b\sin v_\textrm{self-obs} + t\cos i - b\sin v_\textrm{limit}(u_\textrm{t},i) \textrm{ ,}
	\end{aligned}
\end{equation}
which is to be solved for \{$v_\textrm{self-obs}, t, u_\textrm{t}$\}. In order to obtain $\Theta_\mathrm{limit}(i)$, which is the maximum half-opening angle where no reflecting area is directly visible by a generally inclined observer, we need to set $u = 3\pi/2 = u_\mathrm{t} + \pi$, which sets the location in $u$ of the first observed point in the direction away from the observer, if $\Theta$ is increasing from 0, and the corresponding $u_\mathrm{t}$ value for the tangential point towards the observer. And $v_\mathrm{self-obs}(3\pi/2,i) = \pi - \arcsin{(\rho\cot{\Theta_\mathrm{limit}}/b)}$, which is given by the upper shadow boundary for the emission point. Numerical solutions of $\Theta_\mathrm{limit}(i)$ for selected values of $\rho/\rho_\mathrm{in}$ are shown in Fig. \ref{fig:torus_curves} in blue. For a special case of a circular torus $\rho = \rho_\mathrm{c} (\Theta)$ and $v_\mathrm{self-obs}(3\pi/2,i) = \pi - \Theta_\mathrm{limit,c}$. The set of equations (\ref{cond2}) then reduces to
\begin{equation}
\begin{split}
    0 &= \cos \Theta_\mathrm{limit,c} + (\sin v_\mathrm{limit}(\pi/2,i) - \sin \Theta_\mathrm{limit,c}) \,\tan i  \\ & \quad \quad \quad -\frac{2}{\cos \Theta_\mathrm{limit,c}} - \cos v_\mathrm{limit}(\pi/2,i) \textrm{ ,}
\end{split}
\end{equation}
which for $a = b$ in (\ref{vlimit}) has an analytical solution\\
\begin{strip}
\hrule
\vspace{6pt}
\begin{align}
i
=\,
&\arctan_2\,\Biggl\{\!\,
  \frac{
    -\cos\Theta_{\mathrm{limit,c}}
    \;+\;
    2\,\sec\Theta_{\mathrm{limit,c}}
    \;-\;
    \sqrt{\,
      \sin^{4}\!\Theta_{\mathrm{limit,c}}
      \;-\;
      5\,\sin^{2}\!\Theta_{\mathrm{limit,c}}
      \;+\;
      4\,\tan^{2}\!\Theta_{\mathrm{limit,c}}
      \;+\;
      \sin^{2}\!\Theta_{\mathrm{limit,c}}\,
      \cos^{2}\!\Theta_{\mathrm{limit,c}}
    }
  }{
    \sin^{2}\!\Theta_{\mathrm{limit,c}}
    \;+\;
    \cos^{2}\!\Theta_{\mathrm{limit,c}}
    \;+\;
    4\,\sec^{2}\!\Theta_{\mathrm{limit,c}}
    \;-\;
    4
  }\textrm{ ,}
\notag\\[6pt]
&\quad
  -\,\csc\,\!\Theta_{\mathrm{limit,c}}\,
  \Biggl[\;
    \frac{
      \cos^{2}\!\Theta_{\mathrm{limit,c}} + 4\,\sec^{2}\!\Theta_{\mathrm{limit,c}} - 4
    }{
      \sin^{2}\!\Theta_{\mathrm{limit,c}}
      + \cos^{2}\!\Theta_{\mathrm{limit,c}}
      + 4\,\sec^{2}\!\Theta_{\mathrm{limit,c}}
      - 4 
    }-\; 1
\notag\\[6pt]
&\quad
    + \frac{
        \bigl(\cos\!\Theta_{\mathrm{limit,c}}-2\,\sec\,\!\Theta_{\mathrm{limit,c}}\bigr)\,
        \sqrt{\,
          \sin^{4}\!\Theta_{\mathrm{limit,c}}
          \;-\;
          5\,\sin^{2}\!\Theta_{\mathrm{limit,c}}
          \;+\;
          4\,\tan^{2}\!\Theta_{\mathrm{limit,c}}
          \;+\;
          \sin^{2}\!\Theta_{\mathrm{limit,c}}\,
          \cos^{2}\!\Theta_{\mathrm{limit,c}}
        }
      }{
        \sin^{2}\!\Theta_{\mathrm{limit,c}}
        + \cos^{2}\!\Theta_{\mathrm{limit,c}}
        + 4\,\sec^{2}\!\Theta_{\mathrm{limit,c}}
        - 4
      }
     \;
  \Biggr]
\Biggr\}\! \textrm{ .}
\end{align}
\vspace{6pt}
\hrule
\label{circ_limit}
\end{strip}
\hspace{-2pt}Its inverse function $\Theta_\mathrm{limit,c}(i)$ for the circular torus with $\rho_\mathrm{c}$ is also shown in Fig. \ref{fig:torus_curves} as a dashed black line.

\subsection{Cone}

For the cone geometry, we adopt the same numerical scheme in a different parametrization. In the Cartesian coordinate system $\{x,y,z\}$ with base vectors $\Vec{e}_\textrm{x} = (1,0,0)$, $\Vec{e}_\textrm{y} = (0,1,0)$, $\Vec{e}_\textrm{z} = (0,0,1)$ we use
\begin{equation}\label{coordtrafo_cone}
	\begin{aligned}
		x &= \left( \rho_\mathrm{in} + |v| \,\frac{\rho - \rho_\mathrm{in}}{\rho}\tan \Theta \right)\cos{u} \\
		y &= \left( \rho_\mathrm{in} + |v| \, \frac{\rho - \rho_\mathrm{in}}{\rho}\tan \Theta \right)\sin{u} \\
            z &= v \textrm{ .}
	\end{aligned}
\end{equation}
We divide the range in $\pi/2 < u \leq 3\pi/2$ in $N_\mathrm{u}$ points in linear binning, while the other half-space $3\pi/2 < u \leq 5\pi/2$ is added symmetrically with the opposite sign of the resulting Stokes parameter $U$. We divide the range in $- \rho /\tan{\Theta} < v < \rho /\tan{\Theta} $, for $B = 1$, or $0 < v < \rho /\tan{\Theta} $, for $B = 0$, in $N_\mathrm{v}$ points in linear binning. Equally to the torus case, we opted for $N_\mathrm{v} = 80$ and $N_\mathrm{u} = 50$ linearly spread points, which produces converging results to an exact solution for a majority of the parameter space explored. In extreme cases of small $\Theta$ and large $i$, we had to use a finer integration grid.
The implicit formula for the cone surface is
\begin{equation}
\Phi = x^2 + y^2 - \left( \rho_\mathrm{in} + |z| \,\frac{\rho - \rho_\mathrm{in}}{\rho} \tan \Theta \right)^2 = 0 \textrm{ ,}
\end{equation}
which is used to calculate the local reflection angles and the local normal vector, similarly to the torus. The contributing local reflecting surface of the double cone is
\begin{equation}
\begin{split}
    A_\textrm{u,v} = \int_{u_1}^{u_2} \int_{v_1}^{v_2} & \left( \rho_\mathrm{in} + |v| \,\frac{\rho - \rho_\mathrm{in}}{\rho}\tan \Theta \right) \, \,  \,   \\
    &\cdot \sqrt{1+\left( \frac{\rho - \rho_\mathrm{in}}{\rho}\tan \Theta \right)^2} \, \mathrm{d}v \, \mathrm{d}u \textrm{ .}
\end{split}
\end{equation}

\begin{figure} 
\centering
\includegraphics[width=1\linewidth]
{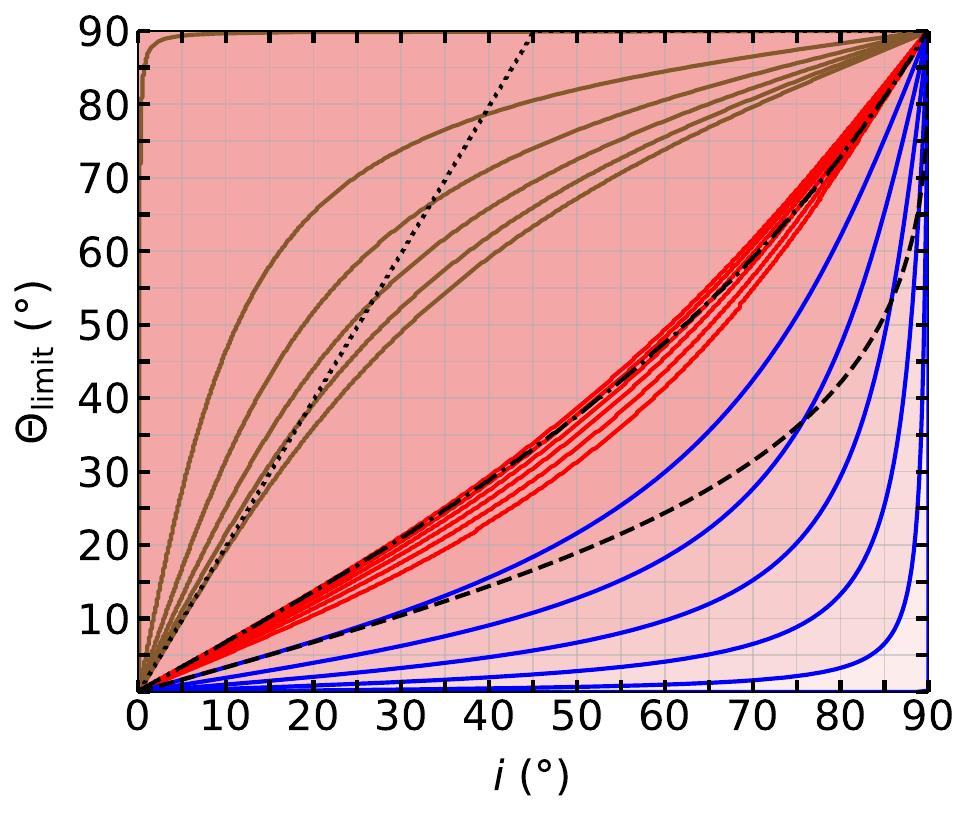}
\caption{\textbf{In blue} we show the $\Theta_\mathrm{limit}(i)$ curves for the elliptical torus, which determine the lower limit in $\Theta$, below which no directly illuminated part of the elliptical torus is directly observable due to self-obscuration. From bottom to top, we show the cases of $\rho/\rho_\mathrm{in} = 1.001,1.2,1.4,1.6,1.8,1.999$, respectively. The dashed black line is the inverse of the function (\ref{circ_limit}), i.e. $\Theta_\mathrm{limit,c}(i)$ for the special case of a circular torus, which effectively spans through all $\rho = \rho_\mathrm{c}(\Theta_\mathrm{limit}) \in (1;2)$ according to equation (\ref{rho_circular}). In our geometrical assumption, such a general limitation to the parameter space due to self-obscuration does not appear for the cone and bowl geometries. \textbf{In red} we show the $\Theta_\mathrm{limit}(i)$ curves for the cone geometry, which determine the lower limit in $\Theta$, below which no part of the double-cone inner surface below the equator (at $z < 0$) is directly observable due to self-obscuration. From bottom to top we show the same cases of $\rho/\rho_\mathrm{in} = 1.001,1.2,1.4,1.6,1.8,1.999$, respectively, and the dash-and-dot line represents the $\rho = \rho_\mathrm{c}(\Theta_\mathrm{limit})$ case according to (\ref{rho_circular}). \textbf{In brown} we show the $\Theta_\mathrm{limit}(i)$ curves for the cone geometry, which determine the upper limit in $\Theta$, above which no part of the double-cone inner surface below the equator (at $z < 0$) is directly observable due to self-obscuration. From top to bottom we show the same cases of $\rho/\rho_\mathrm{in} = 1.001,1.2,1.4,1.6,1.8,1.999$, respectively, and the dotted line represents the $\rho = \rho_\mathrm{c}(\Theta_\mathrm{limit})$ case according to (\ref{rho_circular}). Equivalent red, brown, dot-and-dashed, and dotted curves for the bowl and torus geometries are nearly identical. Note that the axes of this figure nearly overlap with the extreme case of $\rho/\rho_\mathrm{in} = 1.001$ in blue and brown colors.} \label{fig:torus_curves}
\end{figure}

As noted in Section \ref{sec:model}, for the cone geometry, the entire inner surface is illuminated by a central source located at (0,0,0), and for any $\Theta$ and $i$, a non-zero reflecting area is visible by the observer. Therefore, we only have to eliminate those sections of the reflecting surface that are, from the observer's point of view, self-obscured by the double cone itself. The upper half of the structure for $z > 0$ is generating a $u_\mathrm{self-obs}(v,i)$ visibility limitation to the $u$ range due to self-obscuration. Similarly to equations (\ref{cond2}), we formulate
\begin{equation}\label{cond1_cone}
	\begin{aligned}
		0 &= \left( \rho_\mathrm{in} + |v| \,\frac{\rho - \rho_\mathrm{in}}{\rho}\tan \Theta \right)\cos u_\mathrm{self-obs} - \rho_\mathrm{in}\cos u_\textrm{t} \\
		0 &= \left( \rho_\mathrm{in} + |v| \,\frac{\rho - \rho_\mathrm{in}}{\rho}\tan \Theta \right)\sin u_\mathrm{self-obs} + t\sin i - \rho_\mathrm{in}\sin u_\textrm{t} \\
            0 &= v + t\cos i \textrm{ ,}
	\end{aligned}
\end{equation}
and obtain
\begin{equation}
    \begin{split}
        u_\textrm{self-obs} & = \arcsin \left\{ 
        \left[  
            \frac{1}{\rho^2}\left( \rho_\mathrm{in} + |v| \,\frac{\rho - \rho_\mathrm{in}}{\rho}\tan \Theta \right)^2 -1 \right.\right.\\[6pt]
        &\quad\quad \left.\left.+\tan^2 i \left(  \cot\Theta - \frac{v}{\rho}  \right)^2  
        \right] \right. \\[6pt]
        &\quad\quad \left.\cdot \,\frac{\rho}{2\left( \rho_\mathrm{in} + |v| \,\frac{\rho - \rho_\mathrm{in}}{\rho}\tan \Theta \right) \tan i \left(  \frac{v}{\rho} - \cot\Theta  \right)} \right\} \textrm{ .}
    \end{split}
\end{equation}

In addition, for $B = 1$, the equatorial rim is for certain inclinations and half-opening angles covering sections of the lower half of the reflecting structure for $z < 0$, providing additional constraints to the $u$ range at $z < 0$ with $u_\mathrm{self-obs, bottom}(v,i)$. This can be translated as
\begin{equation}\label{cond2_cone}
	\begin{aligned}
		0 &= \left( \rho_\mathrm{in} + |v| \,\frac{\rho - \rho_\mathrm{in}}{\rho}\tan \Theta \right)\cos u_\mathrm{self-obs,bottom} - \rho_\mathrm{in}\cos u_\textrm{t} \\
		0 &= \left( \rho_\mathrm{in} + |v| \,\frac{\rho - \rho_\mathrm{in}}{\rho}\tan \Theta \right)\sin u_\mathrm{self-obs,bottom} + t\sin i - \rho_\mathrm{in}\sin u_\textrm{t} \\
            0 &= v + t\cos i \textrm{ ,}
	\end{aligned}
\end{equation}
which provides
\begin{equation}
    \begin{split}
        u_\textrm{self-obs,bottom} & = \arcsin \left\{ 
        \left[  
            \frac{1}{\rho_\mathrm{in}^2}\left( \rho_\mathrm{in} + |v| \,\frac{\rho - \rho_\mathrm{in}}{\rho}\tan \Theta \right)^2 -1 \right.\right.\\[6pt]
        &\quad\quad \left.\left.+\frac{v^2\tan^2 i}{\rho_\mathrm{in}^2} 
        \right] \frac{\rho_\mathrm{in}^2}{2\left( \rho_\mathrm{in} + |v| \,\frac{\rho - \rho_\mathrm{in}}{\rho}\tan \Theta \right) v \tan i } \right\} \textrm{ .}
    \end{split}
\end{equation}
The lower and upper limit in $\Theta$ per $i$, where the $z < 0$ part of the reflecting inner walls is observable, which is calculated from this condition, is shown in red and brown lines, respectively, in Fig. \ref{fig:torus_curves} for various $\rho/\rho_\mathrm{in}$ values, as well as for the $\rho = \rho_\mathrm{c}$ special case in dash-and-dot and dotted lines, respectively, according to equation (\ref{rho_circular}). Hence, the area between the red and brown curves for each $\rho/\rho_\mathrm{in}$ or between the dot-and-dashed and dotted black curves for $\rho = \rho_\mathrm{c}(\Theta_\mathrm{limit})$ defines the region of the parameter space in $\{\Theta,i\}$ where reflection from the bottom section of the structure affects the net emission for $B=1$. This region is the largest for $\rho \rightarrow 1$, i.e. when the structure approaches a cylinder geometry.

\subsection{Bowl}

For the bowl geometry, we also adopt the same numerical scheme in a corresponding parametrization. In the Cartesian coordinate system $\{x,y,z\}$ with base vectors $\Vec{e}_\textrm{x} = (1,0,0)$, $\Vec{e}_\textrm{y} = (0,1,0)$, $\Vec{e}_\textrm{z} = (0,0,1)$ we use
\begin{equation}\label{coordtrafo_bowl}
	\begin{aligned}
		x &= (\rho_\mathrm{in}+a|\cos{v}|)\cos{u} \\
		y &= (\rho_\mathrm{in}+a|\cos{v}|)\sin{u} \\
            z &=
            \begin{cases}
		 b\, (1 - \sin{v}) \textrm{ ,} & \textrm{if} \ v \in \left[0;\dfrac{\pi}{2}\right] \ , \\
		-b\, (1 + \sin{v}) \textrm{ ,} & \textrm{if} \ v \in \left[\pi;\dfrac{3\pi}{2}\right] \ ,
	\end{cases}
	\end{aligned}
\end{equation}
where $a = \rho - \rho_\mathrm{in}$ and $b = \rho/\tan \Theta$. As for the torus and cone, we divide the range in $\pi/2 < u \leq 3\pi/2$ in $N_\mathrm{u}$ points in linear binning, while the other half-space $3\pi/2 < u \leq 5\pi/2$ is added symmetrically with the opposite sign of the resulting Stokes parameter $U$. We divide the range in $v \in [0;\pi/2] \cup [\pi;3\pi/2] $, for $B = 1$, or $v \in [0;\pi/2]$, for $B = 0$, in $N_\mathrm{v}$ points in linear binning. Equally to the torus and cone cases, we opted for $N_\mathrm{v} = 80$ and $N_\mathrm{u} = 50$ linearly spread points, which produces converging results to an exact solution for a majority of the parameter space explored. In extreme cases of small $\Theta$ and large $i$, we had to use a finer integration grid.
The implicit formula for the bowl surface is
\begin{equation}
\Phi = \left( x^2 + y^2 + z_\mathrm{f}^2\,\frac{a^2}{b^2} + \rho_\mathrm{in}^2 - a^2\right)^2 - 4\,\rho_\mathrm{in}^2\,(x^2+y^2) = 0 \textrm{ ,}
\end{equation}
where
\begin{equation}
        z_\mathrm{f} =
        \begin{cases}
		 2b - z \textrm{ ,} & \textrm{if} \ v \in \left[0;\dfrac{\pi}{2}\right] \ , \\
		-2b - z \textrm{ ,} & \textrm{if} \ v \in \left[\pi;\dfrac{3\pi}{2}\right] \ ,
	\end{cases}
\end{equation}
which is used to calculate the local reflection angles and the local normal vector, similarly to the torus and cone. We approximate the contributing local reflecting surface by the area of a tangent rectangle to the bowl:
\begin{equation}
\begin{split}
    A_\textrm{u,v} & = \int_{u_1}^{u_2} \int_{v_1}^{v_2} (\rho_\mathrm{in}+a|\cos v|) \sqrt{a^2 \sin^2v + b^2\cos^2v} \, \mathrm{d}v \, \mathrm{d}u \\
    & \approx (\rho_\mathrm{in}+\mathrm{sgn}(\cos{v})\,\, a\cos v) \sqrt{a^2 \sin^2v +  b^2\cos^2v}\,  \\
    & \,\,\,\,\,\,\,\,\,\,\,\,\,\,\,\,\,\,\,\,\,\,\,\,\,\,\,\,\,\,\,\,\,\,\,\,\,\,\,\,\,\,\,\,\,\,\,\,\,\,\,\,\,\,\,\, \cdot \, (v_2-v_1)(u_2-u_1) \textrm{ .}
\end{split}
\end{equation}

Also for the bowl geometry, the entire inner surface is illuminated by a central source located at (0,0,0), and for any $\Theta$ and $i$, a non-zero reflecting area is visible by the observer. Thus, we only have to eliminate those sections of the reflecting surface that are, from the observer's point of view, self-obscured by the double bowl itself. The upper bowl of the structure for $z > 0$ is generating the following $u_\mathrm{self-obs}(v,i)$ visibility limitation to the $u$ range due to self-obscuration. Similarly to equations (\ref{cond2}), we formulate
\begin{equation}\label{cond1_bowl}
	\begin{aligned}
		0 &= (\rho_\mathrm{in} + a|\cos v|)\cos u_\mathrm{self-obs} - \rho\cos u_\textrm{t} \\
		0 &= (\rho_\mathrm{in} + a|\cos v|)\sin u_\mathrm{self-obs} + t\sin i - \rho \sin u_\textrm{t} \\
            0 &= 
            \begin{cases}
		 b(1-\sin v) + t\cos i - b \textrm{ ,} & \textrm{if} \ v \in \left[0;\dfrac{\pi}{2}\right] \ , \\
		-b(1+\sin v) + t\cos i - b \textrm{ ,} & \textrm{if} \ v \in \left[\pi;\dfrac{3\pi}{2}\right] \ ,
	\end{cases}
    \end{aligned}
\end{equation}
which yields
\begin{equation}
    u_\textrm{self-obs} = 
    \begin{cases}
    \begin{split}
         \arcsin &\ \left\{\left[ 1 
            - \frac{(\rho_\mathrm{in}+a|\cos{v}|)^2}{\rho^2}   - \frac{\sin^2{v}\,\tan^2 i}{\tan^2\Theta}
        \right]\right.\\
        & \quad \left.
            \cdot \,\frac{\rho\,\tan\Theta }{
            2\,\tan i \,\sin{v} \,\bigl(\rho_\mathrm{in}+a|\cos{v}|\bigr)
        } \right\}
        \textrm{,}\text{ if } v \in \left[0;\tfrac{\pi}{2}\right], 
    \end{split}\\[6pt]
    \begin{split}
        \arcsin & \left\{\left[ 1 
            - \frac{(\rho_\mathrm{in}+a|\cos{v}|)^2}{\rho^2}  - \frac{\sin^2{v}\,\tan^2 i}{\tan^2\Theta} \right] 
            \right.\\
        & \quad \left. \cdot \, \frac{\rho\,\tan\Theta }{
            2\,\tan i \,\sin{v} \,\bigl(\rho_\mathrm{in}+a|\cos{v}|\bigr)
        } \right\}
        \textrm{,}\text{ if } v \in \left[\pi;\tfrac{3\pi}{2}\right].
    \end{split}
    \end{cases}
\end{equation}

In addition, for $B = 1$, the equatorial rim is for certain inclinations and half-opening angles covering sections of the lower bowl of the reflecting structure for $z < 0$, providing additional constraints to the $u$ range at $z < 0$ with $u_\mathrm{self-obs, bottom}(v,i)$. This results in the set of equations
\begin{equation}\label{cond2_bowl}
	\begin{aligned}
		0 &= (\rho_\mathrm{in} + a|\cos v|)\cos u_\mathrm{self-obs, bottom} - \rho_\mathrm{in}\cos u_\textrm{t} \\
		0 &= (\rho_\mathrm{in} + a|\cos v|)\sin u_\mathrm{self-obs, bottom} + t\sin i - \rho_\mathrm{in}\sin u_\textrm{t} \\
            0 &= -\, (1+\sin v) + t\cos i\textrm{ ,}
	\end{aligned}
\end{equation}
with a solution
\begin{equation}
    \begin{split}
        u_\textrm{self-obs,bottom} & = \arcsin \left\{ 
        \left[  
            \rho_\mathrm{in}^2 - (\rho_\mathrm{in}+a|\cos{v}|)^2 \right.\right.\\[6pt]
        &\quad\quad \left.\left.- \,\,\frac{\rho^2(\sin{v}+1)^2\,\tan^2i}{\tan^2\Theta}
        \right] \right.\\[6pt]
        &\quad\quad \left. \cdot \,\,\frac{\tan\Theta }{2\,\rho \,\tan i (\sin{v}+1) \, (\rho_\mathrm{in}+a|\cos{v}| )} \right\} \textrm{ .}
    \end{split}
\end{equation}
The lower and upper limit in $\Theta$ per $i$, where the $z < 0$ part of the reflecting inner walls is observable, which is calculated from this condition, is similar to the torus geometry and cone geometry shown in Fig. \ref{fig:torus_curves}.

\section{Normalization implementation}\label{normalization}

The normalization of the local reflection tables from \cite{Podgorny2022} is provided in Eq. 4 therein. The newly created reflection tables for purely neutral slab are constructed in the same way, and the resulting Stokes parameters are stored also in the units of $[\mathrm{counts}\, \mathrm{cm}^{-2}\,\mathrm{s}^{-1}]$. Because we do not consider a true ionization parameter in the neutral case, we normalize the reflected radiation in such a way that the incident radiation flux follows
\begin{equation}
    F_\mathrm{E}^\mathrm{prim} = N_0\,E^\mathrm{-\Gamma+1} \textrm{ ,}
\end{equation}
and $N_0 = 1$, which requires multiplication of the Stokes parameters $I$, $Q$ and $U$, commonly denoted as $S$, from the raw {\tt STOKES} output:
\begin{equation}
    S_\mathrm{tab}^\mathrm{ref} = \frac{C_1}{N_\mathrm{tot}\Delta \mu_\mathrm{e}^\mathrm{tab}\Delta \Phi_\mathrm{e}^\mathrm{tab}}\,S\textrm{ ,}
\end{equation}
where $N_\mathrm{tot}$ is the number of photons per {\tt STOKES} Monte Carlo simulation ($\geq8.5\times10^{10}$), $\Delta \mu_\mathrm{e}^\mathrm{tab} = 0.05$ and $\Delta \Phi_\mathrm{e}^\mathrm{tab} = 15^\circ$ are the used emission inclination and azimuthal bin sizes, respectively, and
\begin{equation}
    C_1 = \begin{cases}
		\dfrac{(E_\mathrm{max}^{1-\Gamma}-E_\mathrm{min}^{1-\Gamma})}{(1-\Gamma)} \ , & \textrm{if }\,\Gamma \neq 1 \ , \\
		 \ln\left( \dfrac{E_\mathrm{max}}{E_\mathrm{min}}\right)   \ , & \textrm{if }\,\Gamma = 1 \ .
	\end{cases}
\end{equation}

We numerically integrate across the inner walls of the reflecting structure the interpolated Stokes parameters ${\Bar{S}}_\textrm{u,v}^\mathrm{ref}$ from the local reflection tables in \{$u,v$\} that fulfill the shadow and visibility conditions, using the following formula for reflected radiation
\begin{equation}
    {S}^\textrm{ref}(i, \Theta, \rho, \rho_\mathrm{in}, B, \xi_0, \beta,p_0, \Psi_0, \Gamma, D; E) = \frac{\sum_\textrm{u,v} \, A_\textrm{u,v} \,  \, {\Bar{S}}_\textrm{u,v}^\mathrm{ref}}{\left(\frac{D}{\rho_\mathrm{in}}\right)^2}  \textrm{ ,}
\end{equation}
and convert the summed result to the desired units. If the purely neutral case is assumed, we scale each locally interpolated table by an extra factor
\begin{equation}
    N_\mathrm{0,loc}(u,v) = \frac{\xi(u,v)\,n_\mathrm{H,loc}(u,v)}{4\pi C_2}\textrm{ ,}
\end{equation}
where
\begin{equation}
    C_2 = \begin{cases}
		\dfrac{(E_\mathrm{max}^{2-\Gamma}-E_\mathrm{min}^{2-\Gamma})}{(2-\Gamma)} \ , & \textrm{if }\,\Gamma \neq 2 \ , \\
		 \ln\left( \dfrac{E_\mathrm{max}}{E_\mathrm{min}}\right)   \ , & \textrm{if }\,\Gamma = 2 \ ,
	\end{cases}
\end{equation}
and $\xi(u,v)$ is set according to (\ref{xidef}) and $n_\mathrm{H,loc}(u,v)$ according to (\ref{dendef}), so that the incident radiation corresponds to the true locally impinging flux. If the partially ionized case is assumed, each table output is locally interpolated additionally in $\xi(u,v)$ according to (\ref{xidef}) with an additional $\xi(u,v)/\xi_\mathrm{lim}$ normalization factor, if a value outside of the ionization parameter range of the reflection tables is needed, while the table values for the lowest or highest available $\xi_\mathrm{lim}$ are chosen in such cases. We note an extra integration factor of $\mu_\textrm{e}(u,v)$ that is required when using the Chandrasekhar's formulae for local reflection, but that is not required for the local reflection tables. This was not taken into account in \cite{Podgorny2024}, but it has a small impact on the numerical values presented therein and it does not affect any conclusions presented therein.

The primary radiation flux is normalized as
\begin{equation}\label{prim_norm}
    {F}^\textrm{prim}_\mathrm{E}(i, \xi_0, n_\mathrm{H},\Gamma, D, \rho_\mathrm{in}; E) = \frac{\xi_0n_\mathrm{H,0}\rho_\mathrm{in}^2C_3}{4\pi D^2}  \, \frac{E\,a_0(\cos{i})}{E_\mathrm{h}(E\,)-E_\mathrm{l}(E)} \textrm{ ,}
\end{equation}
where $a(\mu)$ is the anisotropy factor defined in Section \ref{sec:model}, $E_\mathrm{h}$ and $E_\mathrm{l}$ are the upper and lower energy bin edges, respectively, and
\begin{equation}\label{norm_factor}
	C_3 = 
	\begin{cases}
		\dfrac{(E_\mathrm{h}^{1-\Gamma}-E_\mathrm{l}^{1-\Gamma})(2-\Gamma)}{(1-\Gamma)(E_\mathrm{max}^{2-\Gamma}-E_\mathrm{min}^{2-\Gamma})} \ , & \textrm{if }\,\Gamma \notin \{1,2\} \ , \\
		 \ln\left( \dfrac{E_\mathrm{h}}{E_\mathrm{l}}\right) \dfrac{2-\Gamma}{E_\mathrm{max}^{2-\Gamma}-E_\mathrm{min}^{2-\Gamma}}  \ , & \textrm{if }\,\Gamma = 1 \ , \\
		 \dfrac{E_\mathrm{h}^{1-\Gamma}-E_\mathrm{l}^{1-\Gamma}}{1-\Gamma} \ln^{-1}\left( \dfrac{E_\mathrm{max}}{E_\mathrm{min}}\right)  \ , & \textrm{if }\,\Gamma = 2 \ .
	\end{cases}
\end{equation}
The corresponding Stokes parameters $Q$ and $U$ for the primary radiation are then inferred from (\ref{prim_norm}) and from the prescribed primary polarization (an)isotropic law in each considered case.

\end{appendix}

\end{document}